\documentclass{article}
\usepackage[T1]{fontenc}
\usepackage{iclr2027_conference,times}

\usepackage{amsmath,amsfonts,bm}

\def\eqref#1{equation~\ref{#1}}

\def\1{\bm{1}}

\DeclareMathAlphabet{\mathsfit}{\encodingdefault}{\sfdefault}{m}{sl}
\SetMathAlphabet{\mathsfit}{bold}{\encodingdefault}{\sfdefault}{bx}{n}

\usepackage{booktabs}
\usepackage{microtype}
\usepackage{hyperref}
\usepackage{url}
\usepackage{multirow}
\usepackage{graphicx}

\usepackage{xcolor}
\usepackage{amssymb}
\usepackage{subcaption}
\usepackage{algorithm}
\usepackage{algpseudocode}
\usepackage[section]{placeins}

\title{Elucidating the Space of Enzymatic Reaction: A Unified Benchmark and Pretrained Model}

\author{
\textbf{
Yutong Hu$^{1,2,\dagger}$,
Tianming Huang$^{1,2,\dagger}$,
Yanbo Zhao$^{1}$,
Qiongyu Zhang$^{3}$,
Shixiang Tang$^{2}$
} \\
\textbf{
Lei Bai$^{2}$,
Ziyi Zhou$^{1}$\thanks{Corresponding authors: Ziyi Zhou, Pan Tan, and Liang Hong.},
Liang Hong$^{1,2}$\footnotemark[1],
Pan Tan$^{1,2}$\footnotemark[1]
} \\
\vspace{0.5em}
$^{1}$Shanghai Jiao Tong University \\
$^{2}$Shanghai Artificial Intelligence Laboratory \\
$^{3}$The University of Sydney \\
\vspace{0.5em}
\texttt{huyutong2020@sjtu.edu.cn},
\texttt{1780693545@sjtu.edu.cn} \\
\texttt{sjtu1669165@sjtu.edu.cn},
\texttt{qzha0437@uni.sydney.edu.au} \\
\texttt{tangshixiang@pjlab.org.cn},
\texttt{bailei@pjlab.org.cn} \\
\texttt{zy-zhou@sjtu.edu.cn},
\texttt{tpan1039@gmail.com} \\
\texttt{hongl3liang@sjtu.edu.cn}
}

\iclrfinalcopy

\usepackage{xspace}
\newcommand{\model}{VenusRX\xspace}
\newcommand{\benchmark}{VenusRX-Bench}

\begin{document}
\maketitle

\begingroup
\renewcommand{\thefootnote}{$\dagger$}
\footnotetext{These authors contributed equally to this work.}
\endgroup

\begin{abstract}
Existing reaction models primarily learn molecular transformations, whereas enzymatic reactions depend jointly on molecular structure and catalytic function. We formulate this problem as learning an \emph{enzymatic reaction space} linking reactants, products, and Enzyme Commission (EC) annotations. To characterize this space, we introduce \benchmark, a unified benchmark for forward reaction prediction, single-step retrosynthesis, and EC-number prediction. \benchmark~integrates reactions from multiple biochemical databases with standardized curation, leakage-controlled splits, and consistent evaluation. Benchmarking representative chemical and enzymatic models reveals a clear chemical-to-enzymatic domain gap, driven by limited domain data, catalytic-context dependency, and the difficulty of modeling large biomolecular structures.

To bridge this gap, we develop \model, a unified T5-style sequence-to-sequence model for enzymatic reactions. \model~jointly learns forward prediction, retrosynthesis, and reaction reconstruction, with two-stage training on millions of template-expanded reactions followed by real biochemical reactions. In addition, optional EC conditioning incorporates catalytic context, while Molecule Library-Constrained Decoding improves the generation of complex biomolecules. Across benchmark tasks and challenging generalization splits, \model~achieves the best or competitive performance on most evaluated settings over representative chemical and enzymatic baselines. Moreover, EC information consistently improves reaction prediction, while learned reaction representations support accurate EC prediction, revealing a bidirectional relationship between reaction structure and catalytic function. Together, \benchmark~and \model~provide a unified framework for elucidating and modeling enzymatic reaction space.

\end{abstract}

\section{Introduction}
\label{sec:introduction}

Enzymes are remarkable biocatalysts capable of accelerating complex chemical transformations with extraordinary regio- and stereoselectivity under mild physiological conditions. These superior catalytic properties position enzymatic reactions as a sustainable alternative to traditional chemical synthesis, driving transformative applications in synthetic biology, pharmaceutical manufacturing, and green chemistry~\citep{turner2013biocatalysis,sheldon2018role,nielsen2016engineering}.
With the burgeoning demand for enzymatic synthesis, deciphering and modeling the underlying principles of enzymatic reactivity has emerged as a fundamental endeavor. In the pursuit of automated biocatalytic pathway design, data-driven approaches capable of predicting reaction outcomes, proposing biosynthetic routes, and inferring catalytic functions directly from molecular representations hold great potential~\citep{delepine2018retropath2,finnigan2021retrobiocat}.

Currently, machine learning and deep learning methods have achieved substantial progress in conventional organic reaction modeling, including reaction outcome prediction, retrosynthesis, and reaction classification. This progress has been largely driven by the availability of large-scale reaction datasets extracted from chemical patents~\citep{lowe2012extraction,kearnes2021open}, which have enabled efficient benchmarking~\citep{wan2022retroformer} and pretraining~\citep{schwaller2021rxnfp,irwin2022chemformer} of AI models. However, for enzymatic reaction, the scarcity of high-quality, large-scale datasets and benchmarks has become a major bottleneck. First, reported enzymatic reactions are limited in quantity and scattered across heterogeneous databases, requiring extensive standardization, correction, and curation~\citep{heid2023enzymemap}. Second, existing studies for modeling enzymatic reactions vary in data sources, preprocessing pipelines, task definitions, and data splitting strategies~\citep{probst2022biocatalysed,zheng2022bionavi,heid2023enzymemap}, making their reported performances difficult to compare directly.

\begin{figure*}[t]
    \centering
    \includegraphics[width=\textwidth]{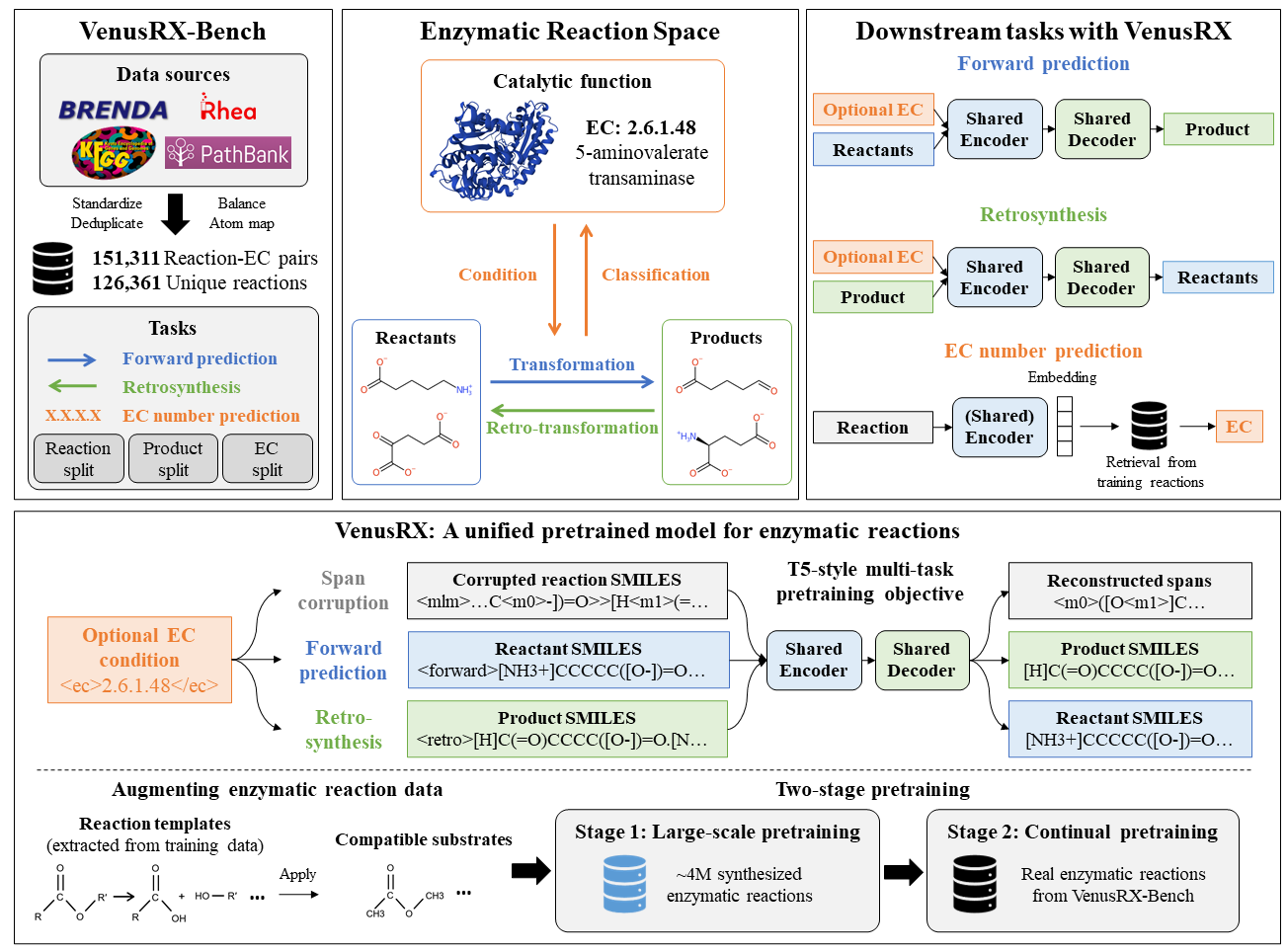}

    \caption{
    \textbf{Overview of the enzymatic reaction space and our framework.}
    \benchmark~evaluates the relationships among reactants, products,
    and catalytic functions, while \model~learns a shared model of
    this space through domain-specific pretraining.
    }
    \label{fig:overview}
\end{figure*}

Moreover, enzymatic reactions differ from organic reactions in a fundamental way: the enzyme catalyst is not merely a rate accelerator but an active determinant of reaction outcome. Organic reaction outcomes are primarily governed by substrate functional group chemistry, yielding relatively consistent pathways across different conditions. In contrast, the high functional diversity of enzymes introduces additional dimensions into enzymatic reactions—including substrate specificity, stereoselectivity, and cofactor dependence—such that reaction outcomes depend jointly on both substrate and enzyme. Since existing organic reaction models are not designed to incorporate enzymatic features, they are insufficient to capture the complexity of enzymatic transformations. In this work, we formulate the \textbf{enzymatic reaction space} as the joint relationships among \emph{reactants}, \emph{products}, and \emph{catalytic functions} (Figure~\ref{fig:overview}). Specifically, we employ Enzyme Commission (EC) numbers as a proxy for catalytic function, as their hierarchical taxonomy systematically encodes enzyme specificity while reaction-EC pairs are more abundant in existing databases than reaction-enzyme pairs.

To chart the enzymatic reaction space, we first construct a comprehensive benchmark named \textbf{\benchmark}. It comprises 126,361 unique enzymatic reactions curated from multiple biochemical databases, all annotated with EC numbers and atom-atom mapping. To the best of our knowledge, it represents the largest curated dataset of enzymatic reactions to date. \benchmark~ covers three core tasks: forward reaction prediction, single-step retrosynthesis, and EC-number prediction. It employs leakage-controlled data splits to evaluate out-of-distribution generalization across distinct biochemical dimensions. Our evaluation reveals that directly transferring organic reaction models to enzymatic reactions yields suboptimal performance, primarily due to the organic-to-enzymatic domain gap: (1) the presence of enzymatic catalysis, and (2) the larger molecular size of biomolecules. Given the limited scale of enzymatic reaction data, this gap cannot be easily bridged by simple supervised learning.

Accordingly, we propose \textbf{\model}, a unified pretrained language model for enzymatic reactions. \model adopts an encoder-decoder architecture and is pretrained under a T5-style~\citep{raffel2020exploring} multi-task objective on enzymatic reaction SMILES and EC numbers, encompassing span corruption, forward reaction prediction, and retrosynthesis. To address the scarcity of enzymatic reaction data, we generate million-scale synthetic enzymatic reactions based on extracted reaction templates and conduct two-stage pretraining on both synthetic and real enzymatic reactions. To incorporate catalytic function context, we randomly append EC prompts to the input during pretraining, enabling the model to optionally condition its predictions on EC numbers at inference time. To improve the generation of large biomolecules, we further design a trie-based constrained sampling strategy that effectively reduces invalid outputs. Experimental results demonstrate that \model achieves the best or competitive performance on most evaluated settings compared to both organic-chemistry and biochemical baselines, and that each of the above designs contributes critically to model performance.

\section{\benchmark}

\label{sec:benchmark}

Biochemical reaction databases differ in their formats, identifiers, and structural coverage.
Study-specific tasks and processing procedures further limit direct comparison of reported results.
\benchmark~addresses these differences through a common reaction collection, standardized prediction tasks, and shared evaluation rules.

\subsection{Data Construction and Splits}
\label{sec:bench-construction}
\label{sec:bench-splits}

\paragraph{Data Construction}
We integrate EC-annotated reactions from BRENDA, PathBank, Rhea, and KEGG into a common representation while retaining source identifiers for traceability.
Records are deduplicated by normalized reaction text and EC annotation, preserving associations between the same reaction and different catalytic annotations.
We then resolve molecular structures, use EnzymeMap~\citep{heid2023enzymemap} for reaction balancing where possible, and generate atom mappings with RXNMapper~\citep{RXNMapper}.
Following ECReact~\citep{probst2022biocatalysed}, product-side curation retains a single principal product and excludes records without reactants.

Table~\ref{tab:dataset_scale} compares the integrated collection with existing biochemical reaction datasets.
\benchmark~has the largest reported reaction count among the listed datasets and provides both EC annotations and atom mappings.

\paragraph{Splits}
We employ three split strategies for training, validation, and test sets.
In each split, the grouping unit is disjoint across the training, validation, and test sets:
\begin{itemize}
    \setlength{\itemsep}{2pt}
    \setlength{\parskip}{0pt}
    \item \textbf{Reactant split.}
    Records are grouped by canonical reactant combinations to evaluate generalization to unseen combinations.
    \item \textbf{Product split.}
    Records are grouped by canonical principal products to evaluate generalization to unseen target products.
    \item \textbf{EC split.}
    Records are grouped by their original complete four-level EC annotations to evaluate generalization to unseen catalytic activities.
\end{itemize}
Product disjointness applies to the product side, while parent EC classes may overlap under the EC split.
Detailed curation procedures and additional grouping constraints are provided in Appendix~\ref{app:benchmark}.

\subsection{Tasks and Evaluation}
\label{sec:bench-tasks}
\label{sec:bench-evaluation}

Each benchmark example is represented as $(R,P,e)$, where $R$ denotes
the reactants, $P$ the retained principal product, and $e$ the associated
EC annotation. We define three tasks:
\[
\begin{aligned}
\text{Forward prediction:}\quad
& R \xrightarrow{\,e\ \text{(optional)}\,} P, \\
\text{Single-step retrosynthesis:}\quad
& P \xrightarrow{\,e\ \text{(optional)}\,} R, \\
\text{EC-number prediction:}\quad
& (R,P) \longmapsto e.
\end{aligned}
\]

For forward prediction and retrosynthesis, we report multi-reference
Top-$k$ exact-match accuracy for $k\in\{1,3,5,10\}$.
When multiple valid targets are associated with the same input in the
test set, a prediction is counted as correct if any of the top-$k$
candidates matches any recorded reference.
Results with and without EC conditioning are reported separately.

For EC-number prediction, we evaluate all four hierarchy levels
(EC1--EC4) using top-1 accuracy, weighted F1, and macro F1.
At level $\ell$, the target consists of the first $\ell$ fields of the
EC annotation. The EC label is used only as the prediction target and is
never provided as input.


\subsection{Dataset Property}
\label{sec:bench-analysis}
Table~\ref{tab:molecule_statistics} compares \benchmark~with USPTO50K.
\benchmark~contains more reactant-side heavy atoms on average.
The largest individual reactant is also larger on average.
Thus, the increase reflects larger molecular structures, not simply more reactant components.

\begin{table*}[t]
\centering
\scriptsize

\begin{subtable}[t]{0.48\textwidth}
\centering
\caption{Dataset comparison.}
\label{tab:dataset_scale}
\setlength{\tabcolsep}{4pt}

\begin{tabular}{lrcc}
\toprule
\textbf{Dataset}
& \textbf{\# Rxns}
& \textbf{EC}
& \textbf{Mapping} \\
\midrule
\benchmark~(ours)
& \textbf{126,361}& \checkmark
& \checkmark \\

EnzymeMap~\citep{heid2023enzymemap}
& 47,974
& \checkmark
& \checkmark \\

ECReact~\citep{probst2022biocatalysed}
& 62,222
& \checkmark
& $\times$ \\

CARE~\citep{yang2024care}
& 48,392
& \checkmark
& $\times$ \\

EnzymeCage~\citep{liu2026enzymecage}
& 29,608
& \checkmark
& \checkmark \\

BioNavi-NP~\citep{zheng2022bionavi}
& 33,710
& $\times$
& $\times$ \\
\bottomrule
\end{tabular}

\vspace{2pt}
\begin{minipage}{0.96\linewidth}
\tiny
\end{minipage}

\end{subtable}
\hfill
\begin{subtable}[t]{0.48\textwidth}
\centering
\caption{Molecular statistics.}
\label{tab:molecule_statistics}
\setlength{\tabcolsep}{3.2pt}

\begin{tabular}{llrrr}
\toprule
\textbf{Metric}
& \textbf{Dataset}
& \textbf{Mean}
& \textbf{Med.}
& \textbf{Max} \\
\midrule

\multirow{2}{*}{Sum of heavy atoms}& \benchmark
& \textbf{48.78}
& \textbf{44}
& \textbf{266} \\
& USPTO50K
& 28.29
& 28
& 97 \\

\midrule

\multirow{2}{*}{SMILES length}& \benchmark
& \textbf{234.56}
& \textbf{208}
& \textbf{844} \\
& USPTO50K
& 189.83
& 186
& 707 \\

\midrule

\multirow{2}{*}{molecule count}& \benchmark
& \textbf{2.11}
& 2
& \textbf{5} \\
& USPTO50K
& 1.71
& 2
& 3 \\

\midrule

\multirow{2}{*}{Max reactant heavy atoms}& \benchmark
& \textbf{36.23}
& \textbf{33}
& \textbf{265} \\
& USPTO50K
& 21.87
& 21
& 97 \\

\bottomrule
\end{tabular}

\end{subtable}

\caption{
Comparison of \benchmark~with existing reaction datasets.
(a) \benchmark~provides the largest collection of unique enzymatic
reactions together with EC annotations and atom--atom mappings.
(b) Compared with USPTO50K, \benchmark~contains larger and more
complex reactant structures.All metrics are reactant-side.
}
\label{tab:dataset_comparison}

\end{table*}

These structural differences affect reactant encoding in forward prediction and target generation in retrosynthesis.
In particular, longer, multi-component retrosynthetic targets require models to generate larger molecular structures while maintaining valid representations.
This motivates training directly on enzymatic reactions and constraining molecular generation, rather than assuming reliable transfer from conventional chemical reaction data.
The paired structure-only and EC-conditioned settings separately examine how catalytic information guides prediction.
These considerations motivate \model~in Section~\ref{sec:model}; their effects on transfer, prediction accuracy, and generation validity are assessed in the experiments.

%

\providecommand{\BCtodo}[1]{\textit{[To specify: #1]}}


\section{\model}
\label{sec:model}


\model~is a unified model for learning enzymatic reaction space.
It formulates reaction modeling as shared text-to-text learning over forward prediction, retrosynthesis, and reaction reconstruction.
To address the limited amount of experimentally observed enzymatic reactions, we first train the model on a large template-expanded reaction corpus and then continue training on real \benchmark~reactions.
\model~also supports optional Enzyme Commission (EC) conditioning, allowing catalytic information to guide reaction prediction when it is available.
For retrosynthesis and forward prediction, we further study constrained generation to reduce the difficulty of generating large and complex biochemical molecules.



\subsection{Unified Multitask Modeling}
\label{sec:model-architecture}

\model~uses a T5-style encoder--decoder architecture
\citep{raffel2020t5}, with all parameters shared across three
text-to-text objectives:
\[
\begin{aligned}
\text{Forward:}\quad
& \texttt{<forward>}R \rightarrow P,\\
\text{Retrosynthesis:}\quad
& \texttt{<retro>}P \rightarrow R,\\
\text{Span corruption:}\quad
& \texttt{<mlm>}\widetilde{(R\!\gg\!P)}
  \rightarrow \text{masked spans}.
\end{aligned}
\]
The first two objectives learn both directions of an enzymatic
transformation, while span corruption encourages shared reaction
representations from the complete reaction context.

\paragraph{EC conditioning.}
Because molecular structures alone may not uniquely determine the
intended enzymatic transformation, we additionally treat the EC number
as an optional condition.
For EC-annotated reactions, \texttt{<ec\_number>}$e$ is appended to the
input with probability $0.5$ during training.
This allows the same model to operate with or without catalytic
information at inference time.
Detailed task construction and optimization settings are provided in
Appendix~\ref{app:model_details}.

\subsection{Training Procedure}
\label{sec:model-training}

\model~is trained in two stages.
Stage-1 learns general enzymatic reaction patterns from a large
template-expanded reaction corpus.
Stage-2 then continues training on the real reactions in the
corresponding \benchmark~training split, adapting the pretrained model
to the benchmark distribution.

Both stages use the same model architecture and multitask objectives.
For each reaction, we additionally use R-SMILES augmentation
\citep{zhong2022root} and randomly sample one representation during
training.
This exposes the model to multiple sequence views of the same chemical
reaction without changing the number of unique reactions.



\subsubsection{Large-Scale Enzymatic Reaction Pretraining}
\label{sec:model-general-training}

Experimentally observed enzymatic reactions are limited in both scale
and substrate diversity.
To expand the reaction space available for pretraining, we construct a Stage-1 corpus containing about four million reaction
records.

The corpus is built in four steps:

\begin{enumerate}
    \item \textbf{Template extraction.}
    We extract reaction templates from atom-mapped enzymatic reactions.
    Each template captures the local structural transformation around
    the reaction center and can be reused with new substrates.

    \item \textbf{Candidate substrate collection.}
    We assemble a large molecule library from biochemical and natural
    product databases, standardize molecular structures, and remove
    duplicated or unsuitable candidates.

    \item \textbf{Template-based reaction generation.}
    For each template, we identify structurally compatible substrates
    and apply the reaction rule to generate new reactant--product pairs.
    This expands known enzymatic transformations to a broader range of
    substrate contexts.

    \item \textbf{Reaction filtering.}
    Generated reactions are checked for structural validity, chemical
    feasibility, and duplication.
    To prevent data leakage, we further remove reactions overlapping
    with the validation or test set of each benchmark split before
    pretraining.
\end{enumerate}

This process expands the training data along the chemical dimension,
while R-SMILES augmentation provides multiple sequence representations
of each retained reaction.
The two augmentations therefore play complementary roles:
template-based generation increases reaction diversity, whereas
R-SMILES augmentation increases representation diversity.

Full details of molecule sources, template generation, filtering, and
split-specific processing are provided in
Appendix~\ref{app:synthetic_reaction_generation}.

\subsection{Molecule Library-Constrained Decoding}
\label{sec:model-constraints}

Enzymatic reactions in \benchmark~often involve larger and more complex
molecules than conventional organic reaction datasets
(Table~\ref{tab:molecule_statistics}), making autoregressive
generation more prone to invalid outputs.
To address this issue, we introduce \textbf{Molecule
Library-Constrained Decoding (MLCD)}, which restricts molecular
generation to a predefined library of valid molecules.

MLCD builds a prefix tree (trie) over tokenized molecules in the
library.
During beam search, only next tokens that remain consistent with at
least one library molecule are allowed.
For reactions containing multiple molecules, the constraint is applied
independently to each generated molecule.

The trie constrains the molecular search space, while \model~still
scores and selects candidates according to the input reaction context.
Detailed decoding procedures and pseudocode are provided in
Appendix~\ref{app:mlcd}.

\section{Experiments}
\label{sec:experiments}

\subsection{Experimental Setup}
\label{sec:experimental_setup}

We evaluate forward prediction, single-step retrosynthesis, and
EC-number prediction using the tasks, splits, and metrics defined by
\benchmark.
All \model~experiments use the same 25.1M-parameter configuration.
For reaction generation, we report multi-reference Top-$k$ accuracy,
while weighted F1 is the primary metric for EC prediction.
Results with and without EC conditioning are reported separately.
The best result is shown in \textbf{bold} and the second-best result is \underline{underlined}.

\paragraph{Baselines.}
We compare \model~with representative methods for conventional chemical
reaction modeling, enzymatic reaction modeling, and EC prediction.
Trainable baselines are adapted to the corresponding \benchmark~
training splits, while zero-shot methods use their released pretrained
checkpoints.
Baseline-specific training and inference details are provided in the
appendix.

\paragraph{EC prediction.}
We evaluate \model~representations in both frozen and task-specific
settings.
The frozen setting directly uses pretrained reaction representations
for EC retrieval, while the task-specific setting replaces CLAIRE's
reaction features with the same frozen \model~representations and
retrains its downstream predictor.

\paragraph{Leakage control and round-trip evaluation.}
For each benchmark split, we remove reactions overlapping with its
validation or test sets from the Stage-1 corpus before pretraining.
For round-trip evaluation, we use a fixed \model~forward oracle trained
on the complete benchmark data and measure whether retrosynthetic
candidates can recover the target product.
The oracle is used only for evaluation and does not affect
retrosynthetic generation.
Further implementation details are provided in the appendix.

\subsection{Results}
\label{sec:results}

\subsubsection{Generalization across Enzymatic Reaction Space}
\label{sec:reaction_results}

\begin{table*}[t]
\centering
\caption{
Forward reaction prediction results on three data splits.
We report Top-$k$ exact-match accuracy (\%).
Methods marked with $^{*}$ are evaluated using open-source checkpoints without fine-tuning.
}
\label{tab:forward_results}

\scriptsize
\setlength{\tabcolsep}{3.2pt}

\begin{tabular}{lcccccccccccc}
\toprule
\multirow{2}{*}{\textbf{Method}} &
\multicolumn{4}{c}{\textbf{Reactant Split}} &
\multicolumn{4}{c}{\textbf{Product Split}} &
\multicolumn{4}{c}{\textbf{EC Split}} \\
\cmidrule(lr){2-5}
\cmidrule(lr){6-9}
\cmidrule(lr){10-13}
& T1 & T3 & T5 & T10
& T1 & T3 & T5 & T10
& T1 & T3 & T5 & T10 \\
\midrule

\multicolumn{13}{l}{\textit{Zero-shot}} \\

R-SMILES$^{*}$~\citep{zhong2022root}
& 0.000 & 0.711 & 0.979 & 1.841
& 0.000 & 0.683 & 1.354 & 2.313
& 0.000 & 0.324 & 0.923 & 1.594 \\

\midrule

\multicolumn{13}{l}{\textit{Without EC Input}} \\

R-SMILES
& 70.562 & 82.048
& 85.543 & 88.618
& 30.849 & 44.631
& 50.000 & 55.849
& 15.568 & 28.104
& 34.899 & 44.643 \\

Graph2SMILES~\citep{tu2022graph2smiles}
& 68.453 & 80.219 & 82.863 & 84.576
& 34.408 & 44.427 & 48.286 & 51.702
& 19.823 & 29.794
& 33.401 & 37.548 \\

MolTransformer~\citep{schwaller2019molecular}
& 52.796 & 66.344 & 71.307 & 75.221
& 12.776 & 22.471 & 26.330 & 29.051
& 9.396 & 15.808 & 18.456 & 21.560 \\

MEGAN~\citep{sacha2021megan}
& 28.285 & 39.247 & 44.676 & 46.202
& 19.128 & 24.676 & 26.366 & 27.421
& 7.646 & 12.308 & 14.166 & 14.801 \\

\model
& \underline{73.066} & \underline{84.611}
& \underline{88.048} & \underline{90.739}
& \underline{50.959} & \underline{66.275}
& \underline{71.393} & \underline{77.085}
& \underline{31.100} & \underline{45.170}
& \underline{53.236} & \underline{59.947} \\

\model~(MLCD)
& \textbf{77.074} & \textbf{88.350}
& \textbf{91.414} & \textbf{94.210}
& \textbf{58.785} & \textbf{74.413}
& \textbf{79.123} & \textbf{84.552}
& \textbf{39.274} & \textbf{56.328}
& \textbf{64.753} & \textbf{74.161} \\

\midrule

\multicolumn{13}{l}{\textit{With EC Input}} \\

ECReact~\citep{probst2022biocatalysed}
& 61.440 & 74.080
& 78.285 & 81.885
& 37.812 & 51.558
& 55.908 & 59.456
& 26.246 & 40.400
& 45.602 & 51.989 \\

\model
& \underline{75.221} & \underline{86.778}
& \underline{89.725} & \underline{92.474}
& \underline{55.825} & \underline{69.463}
& \underline{74.041} & \underline{78.811}
& \underline{34.564} & \underline{50.240}
& \underline{56.340} & \underline{64.058} \\

\model~(MLCD)
& \textbf{80.056} & \textbf{91.135}
& \textbf{93.977} & \textbf{95.515}
& \textbf{62.596} & \textbf{77.193}
& \textbf{81.556} & \textbf{85.918}
& \textbf{42.917} & \textbf{61.182}
& \textbf{69.128} & \textbf{77.661} \\

\bottomrule
\end{tabular}
\end{table*}

\begin{table*}[t]
\centering
\caption{
Retrosynthesis results on three data splits.
We report Top-$k$ exact-match accuracy (\%).
Methods marked with $^{*}$ are evaluated using open-source checkpoints without fine-tuning.
}
\label{tab:retrosynthesis_results}

\scriptsize
\setlength{\tabcolsep}{3.2pt}

\begin{tabular}{lcccccccccccc}
\toprule
\multirow{2}{*}{\textbf{Method}} &
\multicolumn{4}{c}{\textbf{Reactant Split}} &
\multicolumn{4}{c}{\textbf{Product Split}} &
\multicolumn{4}{c}{\textbf{EC Split}} \\
\cmidrule(lr){2-5}
\cmidrule(lr){6-9}
\cmidrule(lr){10-13}
& T1 & T3 & T5 & T10
& T1 & T3 & T5 & T10
& T1 & T3 & T5 & T10 \\
\midrule

\multicolumn{13}{l}{\textit{Zero-shot}} \\
SciReasoner$^*$ (LLM)~\citep{wang2025scireasoner}
& 1.445 & 1.957 & 2.528 & 3.355
& 0.947 & 1.570 & 2.121 & 2.876
& 0.204 & 0.611 & 0.839 & 1.282 \\

R-SMILES$^*$
& 0.000 & 0.687 & 2.004 & 2.656
& 0.000 & 1.007 & 2.073 & 2.601
& 0.000 & 0.228 & 0.527 & 0.899 \\

RSGPT$^*$~\citep{rsgpt2025}
& 0.058& 0.082& 0.128& 0.373& 0.000& 0.048& 0.132& 0.204& 0.024& 0.048 & 0.192& 0.527\\

\midrule

\multicolumn{13}{l}{\textit{Without EC Input}} \\
Graph2SMILES
& 7.025 & 11.708 & 15.040 & 18.802
& \textbf{27.397} & \underline{34.192}
& \underline{36.925} & 39.022
& 3.320 & 4.626 & 5.693 & 6.388 \\

Retroformer~\citep{wan2022retroformer}
& 6.710 & 12.488 & 16.170 & 22.938
& 9.863 & 12.800 & 15.688 & 18.253
& 1.558 & 2.601 & 3.248 & 4.219 \\

MEGAN
& 5.510 & 9.914 & 12.022 & 15.645
& 10.127 & 14.190 & 17.581 & 19.343
& 0.683 & 1.534 & 2.061 & 2.469 \\

R-SMILES
& 5.359 & 9.425 & 12.127 & 16.146
& 4.147 & 7.299 & 9.084 & 11.314
& 1.174 & 3.667 & 4.986 & 6.364 \\

BioNavi~\citep{zheng2022bionavi}
& 2.901 & 5.440 & 8.889 & 13.292
& 9.767 & 15.484 & 18.552 & 21.632
& 1.007 & 1.642 & 1.930 & 2.900 \\

G2Retro~\citep{guo2023g2retro}
& 1.351 & 2.365 & 2.971 & 5.720
& 2.385 & 5.177 & 7.227 & 9.012
& 0.503 & 0.923 & 1.594 & 1.822 \\

RSGPT
& 2.854 & 7.712 & 10.869 & 16.181
& 8.497 & 15.580 & 18.816 & 23.730
& 1.151 & 2.601 & 3.679 & 6.628 \\

\model
& \underline{8.446} & \underline{17.183}
& \underline{23.777} & \underline{30.603}
& 20.554 & 30.058 & 34.923 & \underline{39.945}
& \underline{4.878} & \underline{8.737}
& \underline{10.487} & \underline{14.274} \\

\model~(MLCD)
& \textbf{9.448} & \textbf{19.070}
& \textbf{26.584} & \textbf{34.355}
& \underline{22.903} & \textbf{34.636}
& \textbf{40.244} & \textbf{46.093}
& \textbf{7.407} & \textbf{14.657}
& \textbf{18.384} & \textbf{23.849} \\

\midrule

\multicolumn{13}{l}{\textit{With EC Input}} \\
ECReact
& 11.149 & 17.568 & 21.226 & 25.757
& 17.282 & 24.521 & 27.049 & 27.457
& 5.837 & 8.689 & 10.606 & 11.086 \\

\model
& \underline{17.230} & \underline{26.398}
& \underline{29.345} & \underline{34.425}
& \underline{28.847} & \underline{34.696}
& \underline{38.267} & \underline{40.568}
& \underline{18.360} & \underline{24.605}
& \underline{27.481} & \underline{30.872} \\

\model~(MLCD)
& \textbf{17.300} & \textbf{27.400}
& \textbf{32.258} & \textbf{38.828}
& \textbf{29.302} & \textbf{39.525}
& \textbf{43.672} & \textbf{48.466}
& \textbf{20.374} & \textbf{27.505}
& \textbf{30.609} & \textbf{34.552} \\

\bottomrule
\end{tabular}
\end{table*}

Tables~\ref{tab:forward_results} and
\ref{tab:retrosynthesis_results} evaluate reaction prediction across the
Reactant, Product, and EC Splits.
For forward prediction, \model achieves consistently strong performance
across all three settings, with particularly clear gains on the Product
and EC Splits.
Similar improvements are observed in retrosynthesis, where \model
generalizes effectively across changes in molecular targets and catalytic
functions.
Molecule Library-Constrained Decoding (MLCD) provides further gains in
both prediction directions.

These results show that domain-specific pretraining transfers across
multiple dimensions of enzymatic reaction space rather than being tied
to a particular reaction distribution.
They also demonstrate the value of explicitly modeling the structural
characteristics of enzymatic reactions during generation.

\subsubsection{Catalytic Function and Reaction Structure Are Coupled}
\label{sec:catalytic_coupling}

\begin{table*}[t]
\centering
\caption{
EC-number prediction on the Reactant and Product splits.
Weighted F1 (\%) is reported at EC levels 1--4.
}
\label{tab:ec_results}
\label{tab:ec_all_weighted_f1}
\scriptsize
\setlength{\tabcolsep}{3.2pt}
\renewcommand{\arraystretch}{1.08}
\resizebox{\textwidth}{!}{%
\begin{tabular}{lcccccccc}
\toprule
\multirow{2}{*}{\textbf{Method}} &
\multicolumn{4}{c}{\textbf{Reactant Split}} &
\multicolumn{4}{c}{\textbf{Product Split}} \\
\cmidrule(lr){2-5}\cmidrule(lr){6-9}
& EC1 & EC2 & EC3 & EC4 & EC1 & EC2 & EC3 & EC4 \\
\midrule

\multicolumn{9}{l}{\textit{Zero-Shot}} \\

crxnfp~\citep{zhang2025crxnfp}
& $92.38$ & $85.73$ & $80.90$ & $46.54$
& $89.15$ & $80.83$ & $74.35$ & $41.90$ \\

DRFP~\citep{probst2022drfp}
& $\underline{92.57}$ &
$\underline{85.75}$ &
$\underline{81.21}$ &
$\underline{49.40}$ &
$\underline{91.48}$ &
$\mathbf{84.10}$ &
$\mathbf{77.84}$ &
$\underline{44.61}$ \\

rxnfp~\citep{schwaller2021rxnfp}
& $92.11$ & $85.59$ & $80.52$ & $47.11$
& $88.88$ & $79.80$ & $73.03$ & $41.09$ \\

\textbf{\model}$^{*}$
& $\mathbf{94.76}$ &
$\mathbf{87.76}$ &
$\mathbf{82.81}$ &
$\mathbf{51.17}$ &
$\mathbf{92.11}$ &
$\underline{83.80}$ &
$\underline{77.68}$ &
$\mathbf{47.27}$ \\

\midrule

\multicolumn{9}{l}{\textit{Task Specific Model}} \\

Theia~\citep{probst2023theia}
& $93.09 \pm 0.28$ &
$87.72 \pm 0.46$ &
$83.68 \pm 0.24$ &
$\mathbf{52.49} \pm 0.26$ &
$92.48 \pm 0.47$ &
$84.86 \pm 0.38$ &
$\underline{79.33} \pm 0.45$ &
$\mathbf{47.44} \pm 0.36$ \\

CLAIRE~\citep{zeng2025claire}
& $94.88 \pm 0.14$ &
$89.36 \pm 0.27$ &
$\underline{84.14} \pm 0.40$ &
$49.49 \pm 0.96$ &
$92.41 \pm 0.62$ &
$86.18 \pm 0.30$ &
$78.86 \pm 0.51$ &
$43.50 \pm 0.60$ \\

BEC-Pred~\citep{qian2024becpred}
& $\underline{96.16} \pm 0.33$ &
$\underline{89.78} \pm 0.14$ &
$81.45 \pm 0.50$ &
$46.79 \pm 0.43$ &
$\underline{94.67} \pm 0.36$ &
$\underline{87.02} \pm 0.99$ &
$76.13 \pm 0.60$ &
$42.14 \pm 0.76$ \\

\textbf{\model}
& $\mathbf{96.45} \pm 0.11$ &
$\mathbf{90.99} \pm 0.16$ &
$\mathbf{86.03} \pm 0.24$ &
$\underline{51.33} \pm 0.59$ &
$\mathbf{94.85} \pm 0.33$ &
$\mathbf{87.40} \pm 0.22$ &
$\mathbf{80.23} \pm 0.64$ &
$\underline{47.02} \pm 0.28$ \\

\bottomrule
\end{tabular}%
}

\vspace{2pt}
\begin{minipage}{\textwidth}
\footnotesize
{\model} denotes frozen \model~representations followed by a task-specific predictor.
Task-specific results are mean $\pm$ std over three seeds.
\end{minipage}
\end{table*}

Comparing the EC-conditioned and structure-only settings in
Tables~\ref{tab:forward_results} and
\ref{tab:retrosynthesis_results} shows that catalytic information
consistently improves reaction prediction.
The benefit is especially clear in retrosynthesis, where EC annotations
provide useful context for identifying plausible transformations from a
target product.

The reverse relationship is also observed.
As shown in Table~\ref{tab:ec_results}, frozen \model reaction
representations encode sufficient catalytic information to support
effective EC prediction, while using the same representations in a
task-specific predictor further improves performance across the EC
hierarchy.

Together, these results reveal a bidirectional relationship between
reaction structure and catalytic function:
catalytic information guides reaction prediction, while reaction
representations recover catalytic information.
This empirical coupling supports our formulation of enzymatic reaction
space as a joint relationship among reactants, products, and catalytic
functions.

\subsubsection{Consistency and Constrained Generation}
\label{sec:generation_analysis}

MLCD consistently improves exact-match performance across forward prediction and retrosynthesis, showing that constraining the molecular output space helps the model recover benchmark targets more reliably.
The validity analysis in Appendix~\ref{app:mlcd_analysis} further shows that MLCD substantially reduces invalid molecular generation, supporting its effectiveness for decoding large and complex biochemical structures.

Round-trip evaluation in Appendix~\ref{sec:round_trip} provides a complementary view of the generated candidate space.
While unconstrained decoding retains a broader set of alternative hypotheses, MLCD focuses generation on molecules supported by the predefined library.
This more targeted search improves recovery of recorded reaction targets and reduces structurally invalid outputs.

Overall, these results show that MLCD provides an effective inference-time strategy for navigating the large molecular output space of enzymatic reactions, improving both generation validity and target recovery.

\subsubsection{Ablation Study}
\label{sec:ablation}

Table~\ref{tab:ablation} examines the main components of \model.
The full two-stage training configuration provides the strongest overall
performance, showing that large-scale reaction pretraining and
benchmark-domain continued training play complementary roles.
Stage-1 provides transferable enzymatic transformation knowledge, while
Stage-2 adapts this knowledge to real benchmark reactions.

The span-corruption objective provides additional benefits, particularly
for retrosynthesis and EC prediction.
This suggests that reconstructive learning strengthens the shared
reaction representation beyond the two directional generation tasks.
Overall, the ablation results support the combination of large-scale
pretraining, domain adaptation, and multitask learning used by \model.

\begin{table*}[t]
\centering
\caption{
Ablation study on the Reactant Split.
Reaction generation is evaluated with EC conditioning and without MLCD;
EC prediction reports zero-shot weighted F1 (\%).
}
\label{tab:ablation}
\scriptsize
\setlength{\tabcolsep}{2.5pt}

\begin{tabular}{lccccccccccccccc}
\toprule
\multirow{2}{*}{\textbf{Variant}} &
\multicolumn{3}{c}{\textbf{Training Objectives}} &
\multirow{2}{*}{\textbf{Stage-1}} &
\multirow{2}{*}{\textbf{Stage-2}} &
\multicolumn{3}{c}{\textbf{Forward (with EC)}} &
\multicolumn{3}{c}{\textbf{Retrosynthesis (with EC)}} &
\multicolumn{4}{c}{\textbf{EC Prediction}} \\
\cmidrule(lr){2-4}
\cmidrule(lr){7-9}
\cmidrule(lr){10-12}
\cmidrule(lr){13-16}
&
\textbf{Fwd.} &
\textbf{Retro.} &
\textbf{MLM} &
&
&
\textbf{Top-1} &
\textbf{Top-5} &
\textbf{Top-10} &
\textbf{Top-1} &
\textbf{Top-5} &
\textbf{Top-10} &
\textbf{EC1} &
\textbf{EC2} &
\textbf{EC3} &
\textbf{EC4} \\
\midrule

Full
& $\checkmark$ & $\checkmark$ & $\checkmark$
& $\checkmark$ & $\checkmark$
& \textbf{75.221} & \textbf{89.725} & \textbf{92.474}
& \textbf{17.230} & \textbf{29.345} & \textbf{34.425}
& \textbf{94.76} & \textbf{87.76} & \textbf{82.81} & \textbf{51.17} \\

w/o MLM
& $\checkmark$ & $\checkmark$ & --
& $\checkmark$ & $\checkmark$
& 75.175 & 89.119 & 92.020
& 17.160 & 27.691 & 32.013
& 92.53 & 85.14 & 80.31 & 49.54 \\

w/o Stage-1
& $\checkmark$ & $\checkmark$ & $\checkmark$
& -- & $\checkmark$
& 47.192 & 71.086 & 77.178
& 7.130 & 17.882 & 22.682
& 89.86 & 80.78 & 75.49 & 45.17 \\

w/o Stage-2
& $\checkmark$ & $\checkmark$ & $\checkmark$
& $\checkmark$ & --
& 35.450 & 54.578 & 60.881
& 6.454 & 14.003 & 17.999
& 94.41 & 87.01 & 82.03 & 49.38 \\

\bottomrule
\end{tabular}
\end{table*}

\section{Related Work}
\label{sec:related_work}

\subsection{Organic Reaction Modeling}

\paragraph{Forward reaction prediction and retrosynthesis.} A dominant line of work casts reaction modeling as SMILES-to-SMILES translation with Transformers. Molecular Transformer pioneered this paradigm for forward prediction~\citep{schwaller2019molecular}, later extended to single-step retrosynthesis~\citep{karpov2019transformer}; Retroformer further adds local attention over reactive topology~\citep{wan2022retroformer}, while R-SMILES aligns input-output roots via atom mapping to ease sequence learning~\citep{zhong2022root}. In parallel, graph-based models encode molecular topology more explicitly: Graph2SMILES uses a permutation-invariant graph encoder with a SMILES decoder~\citep{tu2022graph2smiles}, whereas MEGAN and Graph2Edits model reactions as sequences of graph edits~\citep{sacha2021megan,zhong2023graph2edits}. These approaches provide useful architectures and baselines for our study.

\paragraph{Chemical reaction pretraining.} To learn transferable reaction knowledge, recent work pretrains models on large chemical corpora before adapting them to downstream tasks. On the representation side, RXNFP learns reaction fingerprints via BERT-style masked language modeling for classification and reaction-space mapping~\citep{schwaller2021rxnfp}, while RxnRep uses chemically consistent contrastive pretraining for classification and regression~\citep{wen2022improving}. On the generative side, Chemformer applies BART-style pretraining for forward prediction and retrosynthesis~\citep{irwin2022chemformer}; ReactSeq introduces a reaction description language for retrosynthesis, retrieval, and yield prediction~\citep{xiong2025bridging}; and RSGPT scales generative pretraining to ten-billion synthetic reactions for retrosynthesis planning~\citep{rsgpt2025}. Their strong performance motivates us to develop a pretrained model tailored specifically to enzymatic reactions.

\subsection{Enzymatic Reaction Data and Modeling}

\paragraph{Data resources and benchmarks.}
Existing efforts have constructed enzymatic reaction datasets and associated evaluation protocols. ECREACT aggregates EC-annotated reactions from Rhea, BRENDA, PathBank, and MetaNetX for forward prediction and retrosynthesis~\citep{probst2022biocatalysed}. EnzymeMap provides a pipeline for correcting, validating, and atom-mapping enzymatic reactions, and evaluates the curated data on forward prediction, retrosynthesis, and regioselectivity~\citep{heid2023enzymemap}. ReactZyme instead focuses on bidirectional enzyme-reaction retrieval, assessing generalization under enzyme and reaction similarity splits~\citep{hua2024reactzyme}. CARE further introduces a benchmark for enzyme classification and retrieval, where Task 2 retrieves EC numbers from query reactions~\citep{yang2024care}. While complementary, these resources differ substantially in task definitions, data processing, and curation quality, making them difficult to merge into a unified benchmark.

\paragraph{Bio-retrosynthesis and catalytic function prediction.}
Beyond dataset curation, several methods model enzymatic transformations by incorporating catalytic context into neural networks. Kreutter et al. adapt a Molecular Transformer via transfer learning, conditioning product prediction on textual enzyme descriptions~\citep{kreutter2021predicting}. ECREACT instead encodes EC numbers as class tokens for forward prediction and jointly predicts reactants and EC classes in retrosynthesis~\citep{probst2022biocatalysed}. BioNavi-NP trains Transformers on organic and biosynthetic reactions and couples single-step predictions with multi-step search for natural-product pathway planning~\citep{zheng2022bionavi}. For catalytic function annotation, CLAIRE employs contrastive learning to predict EC numbers for enzymatic reactions~\citep{zeng2025claire}. More recently, Enzyformer applies two-stage molecular and organic-reaction pretraining before fine-tuning for enzymatic retrosynthesis, together with a separate contrastive model for EC prediction~\citep{liu2026enzyformer}. Despite this progress, existing approaches are largely confined to task-specific supervised learning on individual datasets, and a unified pretraining framework with a comprehensive benchmark dedicated to enzymatic reactions remains lacking.

\section{Conclusions and Future Work}
\label{sec:conclusion}

We introduced \benchmark, a unified benchmark for enzymatic reaction modeling across forward prediction, retrosynthesis, and EC-number prediction, with controlled splits that evaluate generalization across reactants, products, and catalytic functions. Based on this benchmark, we developed \model, a unified pretrained encoder--decoder model that combines large-scale template-expanded pretraining, continued training on real enzymatic reactions, optional EC conditioning, and molecule-library constrained decoding. Experiments across multiple tasks and splits show that domain-specific pretraining and catalytic context are important for modeling enzymatic reactions, while reaction representations themselves also encode useful catalytic information.

Several directions remain for future work. First, EC numbers provide only a coarse description of catalytic function; incorporating enzyme sequences, structures, cofactors, and reaction conditions may enable more precise modeling of enzyme-dependent reaction outcomes. Second, the current benchmark focuses on single-step reactions, and extending the framework to multi-step biosynthetic pathway planning would provide a more complete evaluation of biocatalytic synthesis. Finally, constrained decoding currently relies on a predefined molecular library. Developing generation methods that preserve chemical validity while allowing exploration beyond a fixed candidate space may further improve generalization to previously unseen biochemical molecules. 

\section*{AI use statement}

AI was used to provide feedback on experimental design, assist with code implementation and checking, conduct related-work research, and support manuscript drafting. The correctness of the implementation and experimental results was independently verified by the authors. The authors take full responsibility for the final content of this work.

\section*{Reproducibility Statement.}
We provide detailed descriptions of the benchmark construction, data splits, model architecture, training procedure, inference settings, and evaluation protocols in the main text and appendix. Hyperparameters and implementation details necessary to reproduce the reported experiments are documented where applicable. Upon acceptance, we will publicly release the source code, benchmark processing pipeline, evaluation scripts, and model checkpoints to facilitate reproducibility and future research.

\bibliography{references}

\begin{thebibliography}{40}
\providecommand{\natexlab}[1]{#1}
\providecommand{\url}[1]{\texttt{#1}}
\expandafter\ifx\csname urlstyle\endcsname\relax
  \providecommand{\doi}[1]{doi: #1}\else
  \providecommand{\doi}{doi: \begingroup \urlstyle{rm}\Url}\fi

\bibitem[Degen et~al.(2008)Degen, Wegscheid-Gerlach, Zaliani, and
  Rarey]{degen2008art}
J{\"o}rg Degen, Christof Wegscheid-Gerlach, Andrea Zaliani, and Matthias Rarey.
\newblock On the art of compiling and using 'drug-like' chemical fragment
  spaces.
\newblock \emph{ChemMedChem}, 3\penalty0 (10):\penalty0 1503--1507, 2008.
\newblock \doi{10.1002/cmdc.200800178}.

\bibitem[Del{\'e}pine et~al.(2018)Del{\'e}pine, Duigou, Carbonell, and
  Faulon]{delepine2018retropath2}
Baudoin Del{\'e}pine, Thomas Duigou, Pablo Carbonell, and Jean-Loup Faulon.
\newblock Retropath2.0: A retrosynthesis workflow for metabolic engineers.
\newblock \emph{Metabolic Engineering}, 45:\penalty0 158--170, 2018.
\newblock \doi{10.1016/j.ymben.2017.12.002}.

\bibitem[Deng et~al.(2025{\natexlab{b}})Deng, Zhao, Sun, Chen, Wang, Xue, Li,
  Song, Hsieh, Hou, Pan, Alomar, Ji, and Wang]{deng2025rsgpt}
Yafeng Deng, Xinda Zhao, Hanyu Sun, Yu~Chen, Xiaorui Wang, Xi~Xue, Liangning
  Li, Jianfei Song, Chang-Yu Hsieh, Tingjun Hou, Xiandao Pan, Taghrid~Saad
  Alomar, Xiangyang Ji, and Xiaojian Wang.
\newblock {RSGPT}: a generative transformer model for retrosynthesis planning
  pre-trained on ten billion datapoints.
\newblock \emph{Nature Communications}, 16:\penalty0 7012, 2025{\natexlab{b}}.
\newblock \doi{10.1038/s41467-025-62308-6}.

\bibitem[Deng et~al.(2025{\natexlab{a}})Deng, Zhao, Sun, Chen, Wang, Xue, Li,
  Song, Hsieh, Hou, Pan, Alomar, Ji, and Wang]{rsgpt2025}
Yafeng Deng, Xinda Zhao, Hanyu Sun, Yu~Chen, Xiaorui Wang, Xi~Xue, Liangning
  Li, Jianfei Song, Chang-Yu Hsieh, Tingjun Hou, Xiandao Pan, Taghrid~Saad
  Alomar, Xiangyang Ji, and Xiaojian Wang.
\newblock Rsgpt: a generative transformer model for retrosynthesis planning
  pre-trained on ten billion datapoints.
\newblock \emph{Nature Communications}, 16\penalty0 (1):\penalty0 7012,
  2025{\natexlab{a}}.
\newblock \doi{10.1038/s41467-025-62308-6}.
\newblock URL \url{https://doi.org/10.1038/s41467-025-62308-6}.

\bibitem[Finnigan et~al.(2021)Finnigan, Hepworth, Flitsch, and
  Turner]{finnigan2021retrobiocat}
William Finnigan, Lorna~J. Hepworth, Sabine~L. Flitsch, and Nicholas~J. Turner.
\newblock Retrobiocat as a computer-aided synthesis planning tool for
  biocatalytic reactions and cascades.
\newblock \emph{Nature Catalysis}, 4\penalty0 (2):\penalty0 98--104, 2021.
\newblock \doi{10.1038/s41929-020-00556-z}.

\bibitem[Grandjean et~al.(2026)Grandjean, Raffaud, Westerlund, Schwaller,
  Genheden, and Reymond]{RXNMapper}
Yves Grandjean, Sacha Raffaud, Annie~M. Westerlund, Philippe Schwaller, Samuel
  Genheden, and Jean-Louis Reymond.
\newblock Rxnmapperv2: updated validation and evaluation on expanded chemical
  space.
\newblock \emph{ChemRxiv}, 2026\penalty0 (0626), 2026.
\newblock \doi{10.26434/chemrxiv.15005247/v1}.
\newblock URL \url{https://chemrxiv.org/doi/abs/10.26434/chemrxiv.15005247/v1}.

\bibitem[Guo et~al.(2023)Guo, Wu, Liu, Sun, Yang, Yang, Wang, Zhang, Wang, and
  Du]{guo2023g2retro}
Zhongliang Guo, Yan Wu, Ke~Liu, Yuxin Sun, Yujing Yang, Peng Yang, Jun Wang,
  Yidan Zhang, Jian Wang, and Wen Du.
\newblock {G2Retro} as a two-step graph generative models for retrosynthesis
  prediction.
\newblock \emph{Communications Chemistry}, 6:\penalty0 102, 2023.
\newblock \doi{10.1038/s42004-023-00897-3}.

\bibitem[Heid et~al.(2023)Heid, Probst, Green, and Madsen]{heid2023enzymemap}
Esther Heid, Daniel Probst, William~H. Green, and Georg K.~H. Madsen.
\newblock Enzymemap: Curation, validation and data-driven prediction of
  enzymatic reactions.
\newblock \emph{Chemical Science}, 14\penalty0 (48):\penalty0 14229--14242,
  2023.
\newblock \doi{10.1039/D3SC02048G}.

\bibitem[Hua et~al.(2024)Hua, Zhong, Luan, Hong, Wolf, Precup, and
  Zheng]{hua2024reactzyme}
Chenqing Hua, Bozitao Zhong, Sitao Luan, Liang Hong, Guy Wolf, Doina Precup,
  and Shuangjia Zheng.
\newblock {ReactZyme}: A benchmark for enzyme-reaction prediction.
\newblock In \emph{Advances in Neural Information Processing Systems},
  volume~37, pp.\  26415--26442, 2024.
\newblock URL
  \url{https://proceedings.neurips.cc/paper_files/paper/2024/hash/2e68b2367d2e0bc8dd6f0ff86e07c2eb-Abstract-Datasets_and_Benchmarks_Track.html}.

\bibitem[Irwin et~al.(2022)Irwin, Dimitriadis, He, and
  Bjerrum]{irwin2022chemformer}
Ross Irwin, Spyridon Dimitriadis, Jiazhen He, and Esben~Jannik Bjerrum.
\newblock Chemformer: A pre-trained transformer for computational chemistry.
\newblock \emph{Machine Learning: Science and Technology}, 3\penalty0
  (1):\penalty0 015022, 2022.
\newblock \doi{10.1088/2632-2153/ac3ffb}.

\bibitem[Karpov et~al.(2019)Karpov, Godin, and Tetko]{karpov2019transformer}
Pavel Karpov, Guillaume Godin, and Igor~V. Tetko.
\newblock A transformer model for retrosynthesis.
\newblock In \emph{Artificial Neural Networks and Machine Learning -- ICANN
  2019: Workshop and Special Sessions}, volume 11731 of \emph{Lecture Notes in
  Computer Science}, pp.\  817--830, Cham, 2019. Springer.
\newblock ISBN 978-3-030-30492-8.
\newblock \doi{10.1007/978-3-030-30493-5_78}.

\bibitem[Kearnes et~al.(2021)Kearnes, Maser, Wleklinski, Kast, Doyle, Dreher,
  Hawkins, Jensen, and Coley]{kearnes2021open}
Steven~M. Kearnes, Michael~R. Maser, Michael Wleklinski, Anton Kast, Abigail~G.
  Doyle, Spencer~D. Dreher, Joel~M. Hawkins, Klavs~F. Jensen, and Connor~W.
  Coley.
\newblock The open reaction database.
\newblock \emph{Journal of the American Chemical Society}, 143\penalty0
  (45):\penalty0 18820--18826, 2021.
\newblock \doi{10.1021/jacs.1c09820}.
\newblock URL \url{https://doi.org/10.1021/jacs.1c09820}.

\bibitem[Kreutter et~al.(2021)Kreutter, Schwaller, and
  Reymond]{kreutter2021predicting}
David Kreutter, Philippe Schwaller, and Jean-Louis Reymond.
\newblock Predicting enzymatic reactions with a molecular transformer.
\newblock \emph{Chemical Science}, 12\penalty0 (25):\penalty0 8648--8659, 2021.
\newblock \doi{10.1039/D1SC02362D}.

\bibitem[Liu et~al.(2026{\natexlab{a}})Liu, Xu, Zhan, Lin, and
  Siu]{liu2026enzyformer}
Tiantao Liu, Jiangcheng Xu, Xinke Zhan, Shaolong Lin, and Shirley W.~I. Siu.
\newblock {Enzyformer}: a two-stage pretrained model for enzymatic
  retrosynthesis.
\newblock \emph{Journal of Cheminformatics}, 18:\penalty0 35,
  2026{\natexlab{a}}.
\newblock \doi{10.1186/s13321-026-01164-y}.

\bibitem[Liu et~al.(2026{\natexlab{b}})Liu, Hua, Xu, Zeng, Rao, Zhang, others,
  and Zheng]{liu2026enzymecage}
Yong Liu, Chenqing Hua, Minghao Xu, Tao Zeng, Jiahua Rao, Zhaoyang Zhang,
  others, and Shuangjia Zheng.
\newblock A geometric foundation model for enzyme retrieval with evolutionary
  insights.
\newblock \emph{Nature Catalysis}, 9\penalty0 (2):\penalty0 148--160,
  2026{\natexlab{b}}.
\newblock \doi{10.1038/s41929-026-01478-y}.

\bibitem[Lowe(2012)]{lowe2012extraction}
Daniel~Mark Lowe.
\newblock \emph{Extraction of Chemical Structures and Reactions from the
  Literature}.
\newblock PhD thesis, University of Cambridge, 2012.

\bibitem[Nielsen \& Keasling(2016)Nielsen and Keasling]{nielsen2016engineering}
Jens Nielsen and Jay~D. Keasling.
\newblock Engineering cellular metabolism.
\newblock \emph{Cell}, 164\penalty0 (6):\penalty0 1185--1197, 2016.
\newblock \doi{10.1016/j.cell.2016.02.004}.

\bibitem[Probst(2023)]{probst2023theia}
Daniel Probst.
\newblock Explainable prediction of catalysing enzymes from reactions using
  multilayer perceptrons.
\newblock \emph{bioRxiv}, 2023.
\newblock \doi{10.1101/2023.01.28.526009}.

\bibitem[Probst et~al.(2022{\natexlab{a}})Probst, Manica, Nana~Teukam,
  Castrogiovanni, Paratore, and Laino]{probst2022biocatalysed}
Daniel Probst, Matteo Manica, Yves~Gaetan Nana~Teukam, Alessandro
  Castrogiovanni, Federico Paratore, and Teodoro Laino.
\newblock Biocatalysed synthesis planning using data-driven learning.
\newblock \emph{Nature Communications}, 13\penalty0 (1):\penalty0 964,
  2022{\natexlab{a}}.
\newblock \doi{10.1038/s41467-022-28536-w}.

\bibitem[Probst et~al.(2022{\natexlab{b}})Probst, Schwaller, and
  Reymond]{probst2022drfp}
Daniel Probst, Philippe Schwaller, and Jean-Louis Reymond.
\newblock Reaction classification and yield prediction using the differential
  reaction fingerprint {DRFP}.
\newblock \emph{Digital Discovery}, 1\penalty0 (2):\penalty0 91--97,
  2022{\natexlab{b}}.
\newblock \doi{10.1039/D1DD00006C}.

\bibitem[Qian et~al.(2024)Qian, Wang, Kang, et~al.]{qian2024becpred}
Wenjia Qian, Xiaorui Wang, Yu~Kang, et~al.
\newblock A general model for predicting enzyme functions based on enzymatic
  reactions.
\newblock \emph{Journal of Cheminformatics}, 16\penalty0 (1):\penalty0 38,
  2024.
\newblock \doi{10.1186/s13321-024-00829-x}.

\bibitem[Raffel et~al.(2020{\natexlab{a}})Raffel, Shazeer, Roberts, Lee,
  Narang, Matena, Zhou, Li, and Liu]{raffel2020exploring}
Colin Raffel, Noam Shazeer, Adam Roberts, Katherine Lee, Sharan Narang, Michael
  Matena, Yanqi Zhou, Wei Li, and Peter~J. Liu.
\newblock Exploring the limits of transfer learning with a unified text-to-text
  transformer.
\newblock \emph{Journal of Machine Learning Research}, 21\penalty0
  (140):\penalty0 1--67, 2020{\natexlab{a}}.

\bibitem[Raffel et~al.(2020{\natexlab{b}})Raffel, Shazeer, Roberts, Lee,
  Narang, Matena, Zhou, Li, and Liu]{raffel2020t5}
Colin Raffel, Noam Shazeer, Adam Roberts, Katherine Lee, Sharan Narang, Michael
  Matena, Yanqi Zhou, Wei Li, and Peter~J. Liu.
\newblock Exploring the limits of transfer learning with a unified text-to-text
  transformer.
\newblock \emph{Journal of Machine Learning Research}, 21\penalty0
  (140):\penalty0 1--67, 2020{\natexlab{b}}.
\newblock URL \url{https://jmlr.org/papers/v21/20-074.html}.

\bibitem[Sacha et~al.(2021)Sacha, B{\l}a{\.z}, Byrski,
  D{\k{a}}browski-Tuma{\'n}ski, Chromi{\'n}ski, Loska,
  W{\l}odarczyk-Pruszy{\'n}ski, and Jastrz{\k{e}}bski]{sacha2021megan}
Miko{\l}aj Sacha, Miko{\l}aj B{\l}a{\.z}, Piotr Byrski, Pawe{\l}
  D{\k{a}}browski-Tuma{\'n}ski, Miko{\l}aj Chromi{\'n}ski, Rafa{\l} Loska,
  Pawe{\l} W{\l}odarczyk-Pruszy{\'n}ski, and Stanis{\l}aw Jastrz{\k{e}}bski.
\newblock Molecule edit graph attention network: Modeling chemical reactions as
  sequences of graph edits.
\newblock \emph{Journal of Chemical Information and Modeling}, 61\penalty0
  (7):\penalty0 3273--3284, 2021.
\newblock \doi{10.1021/acs.jcim.1c00537}.

\bibitem[Schwaller et~al.(2019)Schwaller, Laino, Gaudin, Bolgar, Hunter, Bekas,
  and Lee]{schwaller2019molecular}
Philippe Schwaller, Teodoro Laino, Th{\'e}ophile Gaudin, Peter Bolgar,
  Christopher~A. Hunter, Costas Bekas, and Alpha~A. Lee.
\newblock Molecular transformer: A model for uncertainty-calibrated chemical
  reaction prediction.
\newblock \emph{ACS Central Science}, 5\penalty0 (9):\penalty0 1572--1583,
  2019.
\newblock \doi{10.1021/acscentsci.9b00576}.

\bibitem[Schwaller et~al.(2021{\natexlab{a}})Schwaller, Hoover, Reymond,
  Strobelt, and Laino]{schwaller2021rxnmapper}
Philippe Schwaller, Benjamin Hoover, Jean-Louis Reymond, Hendrik Strobelt, and
  Teodoro Laino.
\newblock Extraction of organic chemistry grammar from unsupervised learning of
  chemical reactions.
\newblock \emph{Science Advances}, 7\penalty0 (15):\penalty0 eabe4166,
  2021{\natexlab{a}}.
\newblock \doi{10.1126/sciadv.abe4166}.

\bibitem[Schwaller et~al.(2021{\natexlab{b}})Schwaller, Probst, Vaucher, Nair,
  Kreutter, Laino, and Reymond]{schwaller2021rxnfp}
Philippe Schwaller, Daniel Probst, Alain~C. Vaucher, Vishnu~H. Nair, David
  Kreutter, Teodoro Laino, and Jean-Louis Reymond.
\newblock Mapping the space of chemical reactions using attention-based neural
  networks.
\newblock \emph{Nature Machine Intelligence}, 3\penalty0 (2):\penalty0
  144--152, 2021{\natexlab{b}}.
\newblock \doi{10.1038/s42256-020-00284-w}.

\bibitem[Sheldon \& Woodley(2018)Sheldon and Woodley]{sheldon2018role}
Roger~A. Sheldon and John~M. Woodley.
\newblock Role of biocatalysis in sustainable chemistry.
\newblock \emph{Chemical Reviews}, 118\penalty0 (2):\penalty0 801--838, 2018.
\newblock \doi{10.1021/acs.chemrev.7b00203}.

\bibitem[Tu \& Coley(2022)Tu and Coley]{tu2022graph2smiles}
Zhengkai Tu and Connor~W. Coley.
\newblock Permutation invariant graph-to-sequence model for template-free
  retrosynthesis and reaction prediction.
\newblock \emph{Journal of Chemical Information and Modeling}, 62\penalty0
  (15):\penalty0 3503--3513, 2022.
\newblock \doi{10.1021/acs.jcim.2c00321}.

\bibitem[Turner \& O'Reilly(2013)Turner and O'Reilly]{turner2013biocatalysis}
Nicholas~J. Turner and Elaine O'Reilly.
\newblock Biocatalytic retrosynthesis.
\newblock \emph{Nature Chemical Biology}, 9\penalty0 (5):\penalty0 285--288,
  2013.
\newblock \doi{10.1038/nchembio.1235}.

\bibitem[Wan et~al.(2022)Wan, Hsieh, Liao, and Zhang]{wan2022retroformer}
Yue Wan, Chang-Yu Hsieh, Benben Liao, and Shengyu Zhang.
\newblock Retroformer: Pushing the limits of end-to-end retrosynthesis
  transformer.
\newblock In \emph{Proceedings of the 39th International Conference on Machine
  Learning}, volume 162, pp.\  22475--22490. PMLR, 2022.
\newblock URL \url{https://proceedings.mlr.press/v162/wan22a.html}.

\bibitem[Wang et~al.(2025)Wang, Tang, Deng, Xiao, Liu, Wu, Yao, Li, Su, Wang,
  Zhuang, Ren, Fei, Hu, Chen, Zhou, He, Yue, Yin, Wu, Zheng, Zhou, Xu, Ma, Lu,
  Zhang, Song, Torr, Tang, Ma, Ouyang, and Bai]{wang2025scireasoner}
Yizhou Wang, Chen Tang, Han Deng, Jiabei Xiao, Jiaqi Liu, Jianyu Wu, Jun Yao,
  Pengze Li, Encheng Su, Lintao Wang, Guohang Zhuang, Yuchen Ren, Ben Fei,
  Mingmao Hu, Xin Chen, Dongzhan Zhou, Junjun He, Xiangyu Yue, Zhenfei Yin,
  Jiamin Wu, Qihao Zheng, Yuhao Zhou, Huihui Xu, Chenglong Ma, Yan Lu, Wenlong
  Zhang, Chunfeng Song, Philip H.~S. Torr, Shixiang Tang, Xinzhu Ma, Wanli
  Ouyang, and Lei Bai.
\newblock {SciReasoner}: Laying the scientific reasoning ground across
  disciplines.
\newblock \emph{arXiv preprint arXiv:2509.21320}, 2025.
\newblock \doi{10.48550/arXiv.2509.21320}.

\bibitem[Wen et~al.(2022)Wen, Li, Xiang, Zhang, and Kim]{wen2022improving}
Y.~Wen, Z.~Li, X.~Xiang, D.~Zhang, and S.~Kim.
\newblock Improving machine learning performance on small chemical reaction
  data with unsupervised contrastive pretraining.
\newblock \emph{Chemical Science}, 13\penalty0 (5):\penalty0 1448--1459, 2022.
\newblock \doi{10.1039/D1SC05469D}.
\newblock URL \url{https://doi.org/10.1039/D1SC05469D}.

\bibitem[Xiong et~al.(2025)Xiong, Zhang, Wang, Huang, Shi, Xu, Li, Fu, Kong,
  Wang, Xiong, and Zheng]{xiong2025bridging}
Jiacheng Xiong, Wei Zhang, Yinquan Wang, Jiatao Huang, Yuqi Shi, Mingyan Xu,
  Manjia Li, Zunyun Fu, Xiangtai Kong, Yitian Wang, Zhaoping Xiong, and Mingyue
  Zheng.
\newblock Bridging chemistry and artificial intelligence by a reaction
  description language.
\newblock \emph{Nature Machine Intelligence}, 7:\penalty0 782--793, 2025.
\newblock \doi{10.1038/s42256-025-01032-8}.
\newblock URL \url{https://doi.org/10.1038/s42256-025-01032-8}.

\bibitem[Yang et~al.(2024)Yang, Mora, Liu, Wittmann, Anandkumar, Arnold, and
  Yue]{yang2024care}
Jason Yang, Ariane Mora, Shengchao Liu, Bruce~J. Wittmann, Anima Anandkumar,
  Frances~H. Arnold, and Yisong Yue.
\newblock {CARE}: A benchmark suite for the classification and retrieval of
  enzymes.
\newblock In \emph{Advances in Neural Information Processing Systems},
  volume~37, pp.\  3094--3121, 2024.
\newblock \doi{10.52202/079017-0101}.
\newblock URL
  \url{https://proceedings.neurips.cc/paper_files/paper/2024/hash/05a7ad45d75a3082d7a3a70de8743140-Abstract-Datasets_and_Benchmarks_Track.html}.

\bibitem[Zeng et~al.(2025)Zeng, Guo, Jin, and Luo]{zeng2025claire}
Zishuo Zeng, Jin Guo, Jiao Jin, and Xiaozhou Luo.
\newblock {CLAIRE}: a contrastive learning-based predictor for {EC} number of
  chemical reactions.
\newblock \emph{Journal of Cheminformatics}, 17:\penalty0 2, 2025.
\newblock \doi{10.1186/s13321-024-00944-8}.

\bibitem[Zhang et~al.(2025)Zhang, Marti, Probst, Robinson, and
  Fenner]{zhang2025crxnfp}
Kunyang Zhang, Thierry~D. Marti, Silke~I. Probst, Serina~L. Robinson, and
  Kathrin Fenner.
\newblock Enzyme association for environmental biotransformation reactions
  through contrastive learning of reaction center-specific fingerprints.
\newblock \emph{Bioinformatics}, 2025.
\newblock \doi{10.1093/bioinformatics/btag142}.

\bibitem[Zheng et~al.(2022)Zheng, Zeng, Li, Chen, Coley, Yang, Wu,
  et~al.]{zheng2022bionavi}
Shuangjia Zheng, Tao Zeng, Chengtao Li, Binghong Chen, Connor~W. Coley, Yuedong
  Yang, Ruibo Wu, et~al.
\newblock Deep learning driven biosynthetic pathways navigation for natural
  products with {BioNavi-NP}.
\newblock \emph{Nature Communications}, 13\penalty0 (1):\penalty0 3342, 2022.
\newblock \doi{10.1038/s41467-022-30970-9}.

\bibitem[Zhong et~al.(2023)Zhong, Yang, and Chen]{zhong2023graph2edits}
Weihe Zhong, Ziduo Yang, and Calvin Yu-Chian Chen.
\newblock Retrosynthesis prediction using an end-to-end graph generative
  architecture for molecular graph editing.
\newblock \emph{Nature Communications}, 14:\penalty0 3009, 2023.
\newblock \doi{10.1038/s41467-023-38851-5}.

\bibitem[Zhong et~al.(2022)Zhong, Song, Feng, Liu, Jia, Yao, Wu, Hou, and
  Song]{zhong2022root}
Zipeng Zhong, Jie Song, Zunlei Feng, Tiantao Liu, Lingxiang Jia, Shaolun Yao,
  Min Wu, Tingjun Hou, and Mingli Song.
\newblock Root-aligned {SMILES}: a tight representation for chemical reaction
  prediction.
\newblock \emph{Chemical Science}, 13\penalty0 (31):\penalty0 9023--9034, 2022.
\newblock \doi{10.1039/D2SC02763A}.

\end{thebibliography}
\bibliographystyle{iclr2027_conference}

\appendix

\section{Benchmark Construction Details}
\label{app:benchmark}

This section provides the construction details of \benchmark.
The pipeline consists of four stages: multi-source harmonization,
structural processing and deduplication, single-product curation,
and leakage-controlled data partitioning.

\subsection{Multi-source Harmonization}
\label{app:benchmark-harmonization}

We collect enzymatic reactions with EC annotations from
BRENDA, PathBank, Rhea, and KEGG.
Because these databases differ in record format and reaction
representation, all entries are first converted into a common schema
containing the source identifier, EC annotation, reaction text,
substrates, products, and molecular structures when available.

Source-specific representations are normalized before merging.
Positive integer stoichiometric coefficients in PathBank are expanded
into repeated molecular components, Rhea component separators are
converted into the common reaction format, and the balanced
reaction-level structures provided by KEGG are retained directly.
Reaction text is normalized by standardizing arrow notation and
whitespace.
EC annotations are cleaned only at the representation level; missing
EC numbers are removed rather than inferred or completed.

We first deduplicate records using the joint key of normalized reaction
text and EC annotation.
Thus, records describing the same reaction with different EC
annotations remain distinct.
Starting from 310,209 source records with valid EC annotations,
this procedure yields 89,605 reaction--EC records, corresponding to
80,558 unique reaction texts.

\subsection{Structural Processing and Deduplication}
\label{app:benchmark-structure}

For records represented by compound names, molecular structures are
resolved through a shared name-to-SMILES mapping and converted into
structured reactions.
Reaction balancing is then applied where possible.
Records that already provide a balanced reaction-level representation,
such as the retained KEGG reactions, are used directly.

Atom mapping is generated with RXNMapper.
When a balanced reaction candidate is available, mapping is performed
on that candidate; otherwise, a structurally valid reaction may be
mapped as a fallback.
Records for which no usable mapped reaction can be obtained are removed
from subsequent construction.

A reaction can admit multiple mapped candidates.
For candidate mapping $m$, we define the edit score as
\begin{equation}
S(m)
=
n_{\mathrm{remove}}(m)
+
n_{\mathrm{add}}(m)
+
n_{\mathrm{atom}}(m),
\end{equation}
where $n_{\mathrm{remove}}$ and $n_{\mathrm{add}}$ denote the numbers
of removed and added bonds, respectively, and
$n_{\mathrm{atom}}$ denotes the number of atoms whose local
environments change.
Candidates with the minimum score are retained.
If several candidates attain the same minimum score, all tied
candidates are preserved rather than selecting one arbitrarily.

We subsequently canonicalize every molecule while retaining atom-map
information, sort molecular components independently on the two sides
of the reaction, and reconstruct the normalized mapped reaction.
This removes differences caused only by component ordering.
Structure-level deduplication is then performed using the EC annotation
together with the normalized mapped reaction representation.

\subsection{Single-product Curation}
\label{app:benchmark-single-product}

Many biochemical reactions contain multiple products, cofactors, or
small by-products and therefore do not directly define a unique target
for reaction prediction.
Following the single-product curation protocol used by
ECReact~\citep{probst2022biocatalysed}, we construct one principal
product for each retained reaction.

Specifically, we apply the following rules:
\begin{enumerate}
    \item remove products that also occur on the reactant side of the
    same reaction;
    \item for reactions containing multiple products, remove known
    coenzymes and common by-products;
    \item remove product molecules containing fewer than four heavy
    atoms; and
    \item discard reactions that retain zero or multiple products, or
    contain no reactants after curation.
\end{enumerate}

The coenzyme and by-product filtering follows the ECReact curation
lists, including common phosphate- and nucleotide-related species,
NAD-related cofactors, coenzyme A, SAM, flavins, heme,
iron--sulfur clusters, acetate, and pyruvate.
This filtering is applied only when the original reaction contains
multiple products, so that a reaction whose sole product belongs to
one of these categories is not removed solely by this rule.

We additionally require consistency between the name-level,
unmapped-structure, and mapped-structure representations.
The mapped and unmapped versions therefore contain exactly the same
examples in the same order and differ only in the presence of atom-map
indices.
After structural processing, product curation, and final
single-product deduplication, the benchmark contains
\textbf{83,439} reaction examples.

\subsection{Leakage-controlled Data Splits}
\label{app:benchmark-splits}

All three benchmark splits are constructed from the same 83,439
single-product reactions.
Rather than independently assigning individual rows, we first define
canonical structural keys.
Atom maps are removed, each molecule is canonicalized, molecular
components are sorted, and the two reaction sides are represented as
\begin{align}
k_R &= \operatorname{canon}(R), \\
k_P &= \operatorname{canon}(P), \\
k_{\mathrm{rxn}} &= k_R \mathbin{\Vert} \texttt{>>}
                    \mathbin{\Vert} k_P .
\end{align}
We additionally use normalized reaction text and the hierarchical EC
annotation as grouping keys.

Samples sharing a constrained key are grouped into indivisible
connected components.
Each component is assigned wholly to the training, validation, or test
partition, targeting an approximately $80/10/10$ ratio.
This prevents records connected by the corresponding leakage criterion
from appearing in different partitions.

We construct three complementary splits:

\paragraph{Reactant Split.}
Samples sharing the same canonical reactant set or normalized reactant
text are kept in the same partition.
This prevents identical substrate combinations and identical reactant
descriptions from crossing partitions.

\paragraph{Product Split.}
In addition to the Reactant Split constraints, samples sharing the same
canonical product are kept in the same partition.
The test set therefore evaluates prediction on target products that are
not present in the training partition.

\paragraph{EC Split.}
Samples are grouped according to their complete four-level EC
annotation together with the canonical full reaction.
Consequently, an EC class assigned to the validation or test partition
does not occur in the training partition, while identical complete
reactions are also prevented from crossing partitions.

Because complete connected components are assigned jointly, the final
partition sizes can deviate slightly from the nominal $80/10/10$
ratio.
The resulting split sizes are shown in
Table~\ref{tab:benchmark-split-stats}.

\begin{table}[t]
\centering
\caption{
Statistics of the three leakage-controlled benchmark splits.
All splits are constructed from the same 83,439 curated reactions.
}
\label{tab:benchmark-split-stats}
\small
\setlength{\tabcolsep}{5pt}
\begin{tabular}{lrrr}
\toprule
\textbf{Split} & \textbf{Train} & \textbf{Validation} & \textbf{Test} \\
\midrule
Reactant Split & 66,672 & 8,183 & 8,584 \\
Product Split  & 66,751 & 8,344 & 8,344 \\
EC Split       & 66,751 & 8,344 & 8,344 \\
\bottomrule
\end{tabular}
\end{table}

\subsection{Mapped and Unmapped Representations}
\label{app:benchmark-representations}

For each partition, we provide both mapped and unmapped reaction
representations.
The two versions share the same examples, filtering decisions, and row
ordering.
The unmapped representation is used by models that operate directly on
molecular strings, whereas the mapped representation supports methods
that require atom correspondence for reaction-center identification or
graph-edit supervision.
This paired construction ensures that models requiring different
reaction representations are evaluated on exactly the same underlying
examples.
```

\section{EC Prediction}

\subsection{EC Prediction: Implementation Details}
\label{app:ec_prediction_setup}

\paragraph{Data and label spaces.}
Reaction SMILES are standardized with RDKit, while preserving one complete reaction per record and retaining all products.
The supplied validation partitions are used without resplitting.
After label normalization, the training vocabularies contain 7, 67, 230, and 4,074 classes at EC1--EC4 for the Reactant Split, and 7, 67, 231, and 3,943 classes for the product split.
Labels appearing only in validation or test data are not added to the prediction candidates.
Test examples with unseen labels are retained in the reported evaluation.

\paragraph{Pretrained checkpoints.}
The \model~model has 25.1M parameters, hidden size 384, and six layers in each of the encoder and decoder.
EC prediction uses the encoder only.
We use the final stage-two checkpoints after 80,100 and 72,100 steps for the reaction and product splits, respectively, rather than validation-best checkpoints.
Stage-two pretraining uses the corresponding training partition with 20-fold R-SMILES augmentation.
The sampling weights for forward prediction, retrosynthesis, and span corruption are 0.35, 0.35, and 0.30, respectively.
EC prompting is enabled with probability 0.5, and the EC contrastive objective is disabled.
Thus, the downstream zero-shot setting does not imply that EC annotations were absent during pretraining.

\paragraph{Reaction representations and zero-shot retrieval.}
For a reaction $r$, let $x_r^{(v)}$ be its $v$-th complete R-SMILES view with the \texttt{<mlm>} prefix.
The frozen encoder $E_{\theta}$ produces the reaction representation
\begin{equation}
\mathbf{z}(r) = \frac{1}{10}\sum_{v=1}^{10}
\operatorname{MeanPool}\!\left(E_{\theta}(x_r^{(v)})\right).
\label{eq:ec_reaction_representation}
\end{equation}
Pooling first averages token embeddings within a view and then averages the 10 view-level vectors.
The same precomputed vectors are used by \model$^{*}$ and \model.
The 10-view representation extraction is distinct from the 20-fold augmentation used during pretraining.
\model$^{*}$ uses cosine $k$-nearest-neighbor retrieval with labeled training reactions as its reference database.
The value of $k$ is selected on validation data, without using test scores.
DRFP and rxnfp also use cosine $k$-nearest-neighbor retrieval without downstream parameter training, with $k$ selected on validation data.
crxnfp uses its released encoder weights without downstream training and performs 1-nearest-neighbor retrieval against the training set.

\paragraph{CLAIRE and \model.}
CLAIRE uses its original concatenated DRFP and rxnfp features, with DRFP folded to 256 dimensions.
\model replaces these inputs with $\mathbf{z}(r)$ while keeping the \model~encoder frozen.
The original CLAIRE implementation is reused for projection training; only feature construction is replaced.
Both methods use the same threefold permutation-augmented training set and train a separate projection network for each EC level.
The five-layer projection network has hidden width 1,280, output dimension 128, LayerNorm, and dropout 0.1.
Training uses a learning rate of $5\times10^{-4}$, batch size 6,000, and triplet margin 1.0.
Both methods train for a fixed 2,000 epochs and select the checkpoint with the lowest training loss within the final 20\% of epochs.
This checkpoint-selection rule does not use validation scores or validation-based early stopping.
Retrieval uses Euclidean distances in the learned projection space.


\paragraph{Other trained baselines.}
Theia's classifier and CLAIRE's projection network are randomly initialized and trained on the corresponding training partition, rather than initialized from their released task-trained weights.
Theia uses 10,240-bit DRFP features with radius 2 and a single hidden layer of 1,664 units with Tanh activation.
It is trained with Adam, learning rate $10^{-3}$, batch size 64, and an exponential learning-rate decay factor of 0.9.
Multi-label training records are expanded for this baseline.
Its official stopping rule requires the current validation loss to improve by at least $10^{-3}$ relative to the mean of the preceding five epochs, with a maximum of 50 epochs.
BEC-Pred is initialized from the official rxnfp \texttt{bert\_pretrained} weights and trained for 48 epochs, using batch size 8, learning rate $10^{-5}$, maximum sequence length 512, and no warmup.
Where paper descriptions and released code differ, we follow the official code configurations.
CLAIRE is additionally evaluated at EC4, and BEC-Pred at EC1, EC2, and EC4, by adapting the label space while keeping their training configurations unchanged.
BEC-Pred retains its raw-string label indexing; for reaction-split EC4, this produces 4,081 output classes before normalization, compared with 4,074 normalized benchmark labels.

\paragraph{Evaluation and strict-subset analysis.}
All methods are scored with the same \texttt{score\_ec.py} implementation and scoring parameters.
Let $\mathcal{Y}_{i,\ell}$ be the set of reference labels for reaction $i$ at EC level $\ell$, and $\hat{y}_{i,\ell}$ the top-ranked prediction.
The hit-any top-1 accuracy is
\begin{equation}
\operatorname{Acc}_{\ell}
= \frac{1}{N_{\ell}}\sum_{i=1}^{N_{\ell}}
\mathbf{1}\!\left[\hat{y}_{i,\ell}\in\mathcal{Y}_{i,\ell}\right],
\end{equation}
where $N_{\ell}$ is the number of records evaluated at level $\ell$.
Weighted F1 is the primary metric, with macro F1 and top-1 accuracy also reported.
The supplementary strict-subset scores exclude 278 reaction-split test records and 18 product-split test records flagged as input-structure-equivalent to training reactions.
These records remain in the main test evaluation.
Strict-subset filtering uses the \texttt{test\_leaked\_ids\_input\_only} flags in \texttt{leakage.json}.
Theia, CLAIRE, BEC-Pred, and \model are run with seeds 13, 21, and 42, and their results are summarized by the mean and standard deviation.
crxnfp, DRFP, rxnfp, and \model$^{*}$ are evaluated as deterministic methods and have no across-seed standard deviation.

%


\section{\model~Implementation Details}
\label{app:model_details}

\subsection{Architecture and Reaction Representation}
\label{app:model_representation}

\paragraph{Architecture and initialization.}
The T5-style model shares input embeddings across the encoder and decoder and ties the output embedding to them.
Table~\ref{tab:model_architectures} lists the configurations used for reaction-split scaling; Tiny is used for cross-split comparisons.
All configurations use ReLU feed-forward layers, dropout $0.1$, 32 relative-attention buckets, a maximum relative-distance parameter of 128, and layer-normalization epsilon $10^{-6}$.
The relative-distance parameter is not a maximum sequence length.
The decoder starts with padding token ID 0; EOS has ID 1.

\begin{table}[htbp]
\centering
\small
\setlength{\tabcolsep}{4pt}
\caption{\model~configurations. Parameter counts are recorded wrapper totals, including the inactive projection head.}
\label{tab:model_architectures}
\begin{tabular}{@{}lrrrrrr@{}}
\toprule
Model & Enc./Dec. & $d_{\mathrm{model}}$ & $d_{\mathrm{ff}}$ & Heads & $d_{\mathrm{kv}}$ & Parameters \\
\midrule
Micro   & 4/4 & 256 & 1,024 & 4  & 64 & 7,554,816 \\
Compact & 6/6 & 320 & 1,280 & 5  & 64 & 17,495,296 \\
Tiny    & 6/6 & 384 & 1,536 & 6  & 64 & 25,147,648 \\
Medium  & 6/6 & 768 & 3,072 & 12 & 64 & 100,135,168 \\
\bottomrule
\end{tabular}
\end{table}

The wrapper also instantiates a two-layer projection head, $d_{\mathrm{model}}\rightarrow d_{\mathrm{model}}\rightarrow256$, with a tanh activation.
Its contrastive-loss coefficient is zero, so it does not contribute to the sequence-training objective.
For Tiny, this head contains 246,400 parameters and the T5 backbone contains 24,901,248 parameters.
No enzyme-sequence encoder or additional EC-prediction loss is used in this training protocol.

\paragraph{Tokenizer and length limits.}
Both stages and standard inference load the same fixed 302-token vocabulary.
A custom regular expression identifies SMILES tokens, including bracketed atoms, multi-character elements, bonds, branches, ring indices, stereochemical symbols, and \texttt{>{}>}.
If the expression does not cover an entire string, tokenization falls back to individual characters; unseen tokens map to \texttt{<unk>}.
Input whitespace is removed.
Maximum source and target lengths are 1,200 and 768 tokens, including special tokens and EOS.
Truncation forces the last retained token to EOS.
Batches are dynamically right-padded; source padding is attention-masked and target padding is assigned label $-100$ and excluded from the loss.

\paragraph{Molecules, cofactors, and stereochemistry.}
Molecules on the same side are separated by \texttt{.}, and reaction sides by \texttt{>{}>}.
An original \texttt{R>agents>P} record contributes its reactant and product sides; the middle agents field is not modeled.
The reactant side retains all components present in the prepared view, including recorded cofactors.
Forward targets and retrosynthesis inputs use one principal product, whereas span corruption retains the full product side.
Principal-product selection first prefers components not recognized as common cofactors and then maximizes a SMILES-based size heuristic.
The heuristic counts uppercase non-hydrogen letters rather than computing an exact heavy-atom count.
If every component is identified as a cofactor, the highest-scoring component is retained; ties follow input order.
Cofactor matching uses an explicit list and canonical or uncharged forms.

Atom-map identifiers are removed from model inputs and targets.
Existing chirality and bond-direction symbols are retained; this does not imply complete stereochemical annotation of every record.
Forward and retrosynthesis use their respective aligned views and need not be literal string inverses.

\paragraph{Task serialization.}
Table~\ref{tab:model_task_formats} gives the with-EC formats before EOS is appended.
Removing the complete \texttt{<ec\_number>}$e$ suffix gives the without-EC format.
Here, $\mathsf{C}$ replaces masked spans with sentinel tokens while protecting special tokens and the EC suffix.

\begin{table}[htbp]
\centering
\small
\setlength{\tabcolsep}{4pt}
\caption{Task-specific source and target formats. EOS is appended to both sequences.}
\label{tab:model_task_formats}
\begin{tabular}{@{}lp{0.61\linewidth}l@{}}
\toprule
Task & Source & Target \\
\midrule
Forward & \texttt{<forward>}$R$\texttt{>{}><mask\_product><ec\_number>}$e$ & $p^{\star}$ \\
Retro & \texttt{<retro>}$p^{\star}$\texttt{<ec\_number>}$e$ & $R$ \\
Span & $\mathsf{C}($\texttt{<mlm>}$R$\texttt{>{}>}$P$)\texttt{<ec\_number>}$e$ & Masked spans \\
\bottomrule
\end{tabular}
\end{table}

\subsection{Task Sampling and Span Corruption}
\label{app:model_masking}

\paragraph{Sampling and loss reduction.}
Training samples records randomly, with distributed sharding across devices.
For each presentation, the collator samples one available view uniformly by list position and one task with probabilities $(0.35,0.35,0.30)$.
The task, view, EC inclusion, and corruption mask can change between presentations.
Duplicate stored views are retained, so repeated strings can have greater sampling probability.
The training loss is computed for each device-local batch; distributed training averages the resulting gradients across devices.
This is not necessarily a globally token-weighted average when devices have different target-token counts.
EOS and sentinel tokens contribute to the loss, and longer targets contribute more supervised positions.
No additional task-loss weights, label smoothing, or EC contrastive loss are applied.

\paragraph{Corruption procedure.}
The span task uses a forward R-SMILES view of the full reaction, with an original-reaction fallback when no valid view is available.
After tokenization, the source is initially restricted to 1,199 tokens.
Special tokens and all tokens from \texttt{<ec\_number>} onward are protected; molecular tokens, \texttt{.}, and \texttt{>{}>} remain maskable.
For $M>0$ maskable positions, the target number of masked tokens is
\begin{equation}
K=\max\!\left(1,\left\lceil0.15M\right\rceil\right).
\end{equation}
Span lengths are sampled as
\begin{equation}
\ell=\max\!\left(1,\operatorname{round}(z)\right),
\qquad
z\sim\operatorname{Exponential}(\mathrm{rate}=1/3).
\end{equation}
Thus, 3 is the scale of the underlying exponential distribution, not a measured mean length of the final merged spans.
Sampling extends rightward from a selected start and may revisit previously selected positions.
It stops when the token budget is reached, no eligible positions remain, or the number of attempts reaches ten times the sequence length.
Adjacent masked positions are merged, ordered by their original positions, and restricted to at most 100 spans.

Each retained span is replaced with \texttt{<extra\_id\_i>} in the source.
The target concatenates each sentinel and its missing span in source order, followed by EOS.
No additional terminal sentinel is appended after the final span, and the target does not reconstruct the full reaction.
An empty maskable set yields an EOS-only target.
The final source and target obey the 1,200/768-token limits.

Reaction-center-dependent start weights require atom-map identifiers in the selected view.
The bundled R-SMILES generator removes these identifiers, so views produced through that path use base weights.
The configured center/neighbor/base weights of 4/2/1 do not multiply the decoder loss.

\subsection{EC Serialization and Label Handling}
\label{app:model_ec_details}

When available, a label is sampled uniformly from the cleaned, deduplicated, sorted \texttt{ec4} list; otherwise, selection falls back to the record's \texttt{ec\_number} field.
The chosen string is appended as \texttt{<ec\_number>}$e$ before source EOS with probability $0.5$.
The tokenizer represents EC digits and periods using its existing tokens, rather than assigning one token to each EC class.
Missing annotations always yield unconditioned inputs, so the corpus-wide conditioned fraction need not equal $50\%$.
Because the suffix is appended before truncation, it may be partially or entirely removed from an overlength source.

The implementation does not impose a complete numerical-validity check on all four-part labels; strings containing an unspecified component such as \texttt{1.1.1.-} are not categorically excluded.
Stage-two preparation deduplicates canonical substrate/product pairs without EC in the key and retains the first record, rather than necessarily preserving every EC annotation for that pair.
In grouped retrosynthesis inference, the first sorted available EC annotation is selected.
Inference removes an optional \texttt{EC } prefix and uses the first four components; labels with fewer than four components do not produce a suffix.
The conditions come from benchmark annotations, not from a separately predicted EC label.

\subsection{Pretraining Corpus and Split-Specific Filtering}
\label{app:model_corpus}

\paragraph{Common preprocessing.}
The available pipeline starts from merged upstream reaction records and cluster assignments.
It selects the designated cluster-center records, applies RXNMapper~\citep{schwaller2021rxnmapper}, and constructs up to 20 task-specific R-SMILES views with seed 13 and mixed forward mode.
Mapping checks examine the presence of atom maps, uniqueness of map identifiers within each side, and agreement between the two sides' map-ID sets.
These checks are not a complete elemental or charge-balance test.
Records with mapping or R-SMILES failures are retained and can fall back to their original reaction representation.
One stored view contains four task-specific strings; these are not four distinct reactions.
The common pre-filtering corpus contains 4,168,335 records.

\paragraph{Structural overlap detection.}
For each benchmark split, comparisons remove atom maps and sort components within each reaction side.
Five matching levels are used: canonical isomeric SMILES; canonical SMILES without stereochemistry; uncharged SMILES without stereochemistry; the preceding form after removal of small inorganic components; and that form with repeated components collapsed to a set.
The small-component rule removes molecules with at most one heavy atom, or carbon-free molecules with at most five heavy atoms.
Matching considers both reaction directions and supports matching a complete reactant side with an individual product.
The last two levels additionally require matching EC annotations when the pretraining record has an EC annotation; otherwise, matching is structural only.
Uncharging failures retain the original molecule, and there is no explicit tautomer-canonicalization level.
These are overlap-detection rules, not the upstream definition of a unique reaction.

\paragraph{Template-associated exclusion.}
Let $T_{\mathrm{train}}$, $T_{\mathrm{val}}$, and $T_{\mathrm{test}}$ be the template-ID sets collected from upstream records that match the corresponding benchmark partitions.
Define
\begin{equation}
E=(T_{\mathrm{val}}\cup T_{\mathrm{test}})
\setminus T_{\mathrm{train}}.
\label{eq:model_template_exclusion}
\end{equation}
We remove records that either directly match a validation/test reaction or carry any template ID in $E$.
A scalar template field is used when the template-ID list is unavailable.
Templates shared with training matches are not removed merely because they also occur in held-out matches.
This is post-generation filtering based on matched-record provenance, not training-only template extraction.

\begin{table}[htbp]
\centering
\small
\setlength{\tabcolsep}{4pt}
\caption{Split-specific stage-one corpora. Stored views include duplicates and exclude dynamically constructed fallback views. Records are not audited counts of unique synthetic reactions.}
\label{tab:model_pretraining_counts}
\begin{tabular}{@{}lrrrr@{}}
\toprule
Split & Retained records & 20-view records & Zero-view records & Stored views \\
\midrule
Reaction & 4,044,455 & 3,660,831 & 383,624 & 73,216,620 \\
EC       & 3,853,604 & 3,471,243 & 382,361 & 69,424,860 \\
Product  & 4,047,718 & 3,665,111 & 382,607 & 73,302,220 \\
\bottomrule
\end{tabular}
\end{table}

\paragraph{Audit scope and counting.}
The final audits report zero validation and test reaction overlaps under the specified matching procedure for all three splits.
Validation and test structures are used for exclusion and auditing, not for standard model gradient updates.
This does not establish that the upstream substrate pool excludes held-out molecules, or that all pretraining products and scaffolds are disjoint from the benchmark held-out sets.
The reaction-split filter removes 123,880 records: 9,398 direct reaction matches and 114,482 additional template-associated records.
Table~\ref{tab:model_pretraining_counts} distinguishes retained records from stored views.
Unique real/synthetic reaction counts and their intersection remain unaudited; no separate real/synthetic sampling ratio is configured.


\subsection{Optimization and Checkpoint Selection}
\label{app:model_optimization}

Table~\ref{tab:model_optimization} reports the standard Tiny training settings.
Stage two loads the selected stage-one weights but resets optimizer, scheduler, step counter, and early-stopping state.
The training seed is 13; internal-validation and fixed benchmark-validation sampling seeds are 10013 and 20013, respectively.
Stage-two preprocessing deduplicates canonical substrate/product pairs before an internal training/validation subdivision.
These prepared-record counts are not substituted for the benchmark scorer's original test-row denominators.

\begin{table}[htbp]
\centering
\small
\setlength{\tabcolsep}{5pt}
\caption{Standard Tiny optimization settings. Step limits are maximum budgets, not actual training durations.}
\label{tab:model_optimization}
\begin{tabular}{@{}lll@{}}
\toprule
Setting & Stage one & Stage two \\
\midrule
Initialization & Random & Stage-one best weights \\
Optimizer & AdamW & AdamW \\
Adam betas & $(0.9,0.999)$ & $(0.9,0.999)$ \\
Weight decay & 0 & 0 \\
Peak learning rate & $2\times10^{-4}$ & $5\times10^{-5}$ \\
Warmup steps & 1,000 & 1,000 \\
Schedule & Cosine decay to zero & Cosine decay to zero \\
Gradient clipping & Norm 1.0 & Norm 1.0 \\
Per-device batch size & 32 & 32 \\
Devices / accumulation & 8 / 1 & 8 / 1 \\
Nominal global batch & 256 records & 256 records \\
Precision & BF16 autocast & BF16 autocast \\
Maximum steps & 700,000 & 100,000 \\
Validation interval & 16,000 steps & 1,000 steps \\
Early-stopping patience & 5 validation checks & 5 validation checks \\
\bottomrule
\end{tabular}
\end{table}

\paragraph{Internal validation.}
The historical training configuration groups records connected by shared template IDs and assigns groups to training or validation using a seeded stable hash.
Records without template IDs use a reaction-key fallback.
Hash thresholds are $0.01$ in stage one and $0.05$ in stage two; these are not guarantees of exact validation-record fractions.
Validation uses batch size 32 and a cap of 200 batches per validation loader.
Reported loss averages batch-level scalar losses rather than computing one corpus-wide token-weighted mean.
Distributed sampler padding can increase the reported sample count above the number of distinct records.

\paragraph{Checkpoint selection.}
For the standard Tiny runs, checkpoint selection uses internal \texttt{val\_loss}; the fixed benchmark validation set is monitored separately as \texttt{stage2\_val\_loss} in stage two.
For example, the Tiny reaction-split checkpoint selected at step 75,000 records internal loss $0.098622$ and fixed benchmark validation loss $0.116906$.
These are different metrics and must not be interchanged.
Table~\ref{tab:model_training_steps} distinguishes selected and final training steps.
The no-span ablation samples forward/retro tasks with probabilities $0.5/0.5$; the stage-two-only ablation starts from random initialization without stage-one training.
These are single-seed experiments with different realized training budgets, not strictly compute-matched ablations.

\begin{table}[htbp]
\centering
\small
\setlength{\tabcolsep}{5pt}
\caption{Selected checkpoint step / final training step for each run.}
\label{tab:model_training_steps}
\begin{tabular}{@{}lll@{}}
\toprule
Run & Stage one & Stage two \\
\midrule
Micro, reaction & 544,000 / 630,080 & 86,000 / 91,136 \\
Compact, reaction & 400,000 / 488,312 & 72,000 / 77,074 \\
Tiny, reaction & 288,000 / 378,048 & 75,000 / 80,100 \\
Medium, reaction & 240,000 / 330,792 & 38,000 / 43,076 \\
Tiny, EC & 336,000 / 420,196 & 72,000 / 77,000 \\
Tiny, product & 176,000 / 268,022 & 67,000 / 72,100 \\
Tiny, no span corruption & 176,000 / 267,784 & 47,000 / 52,154 \\
Tiny, stage two only & Not applicable & 93,000 / 98,078 \\
\bottomrule
\end{tabular}
\end{table}


\subsection{Inference and Separately Reported Diagnostics}
\label{app:model_inference}

\paragraph{Direct inference.}
Forward and retrosynthesis evaluation use the selected stage-two checkpoint with and without supplied EC annotations.
The configuration requests 20 R-SMILES views, mixed mode, seed 13, beam size 10, and 10 returned candidates per view, retaining at most 10 final candidates.
Source/target limits remain 1,200/768 tokens and inference batch size is 2.
The candidate count is not a top-$k$ sampling parameter.
There is no additional task-specific fine-tuning after stage two; this is not zero-shot evaluation with respect to benchmark training data.

\paragraph{Retrosynthesis fusion.}
Candidates are canonicalized and deduplicated within each view, so a structure contributes at most once per view.
For a candidate $a$, let $\mathcal{V}(a)$ contain views that generate it, and let $r_j(a)$ be its best rank in view $j$.
Define
\begin{equation}
b(a)=\min_{j\in\mathcal{V}(a)}r_j(a),
\qquad
q(a)=\sum_{j\in\mathcal{V}(a)}\frac{1}{r_j(a)}.
\end{equation}
The implementation sorts by decreasing
\begin{equation}
s_{\mathrm{fusion}}(a)
=q(a)-10^{8}\bigl(b(a)-1\bigr).
\end{equation}
This prioritizes the best rank obtained in any view, then the reciprocal-rank sum; remaining ties retain first-occurrence order.
It combines representations of one model, not independent models or averaged sequence log probabilities.

\paragraph{Molecule-library constraints.}
The constrained-decoding diagnostic applies token-level constraints from a split-specific molecule library containing train, validation, and test R-SMILES molecules, preserving their SMILES strings.
It measures performance with access to a benchmark-wide molecular vocabulary and is kept separate from unrestricted held-out evaluation.
Library membership does not establish reaction feasibility or enzyme activity, and this setting is not a guarantee of an accuracy upper bound.

\paragraph{Forward rescoring.}
Forward rescoring uses the forward task of the same Tiny stage-two checkpoint, with length-normalized teacher-forced scoring of the target product given a candidate reactant mixture.
It does not use a separately trained forward model.
Two diagnostics are distinguished: the earlier raw-score setting with $\lambda=0.5$, and a rank-blend setting with
\begin{equation}
s_{\mathrm{blend}}(a)
=-\bigl(r_{\mathrm{fusion}}(a)-1\bigr)
-\lambda\bigl(r_{\mathrm{forward}}(a)-1\bigr),
\end{equation}
where larger scores rank first and the fixed main rank-blend setting is $\lambda=0.25$.
The test-set sweep over $\lambda$ is a sensitivity diagnostic, not validation-based parameter selection.
Neither rescoring variant is assumed to improve every condition.


%

\section{Synthetic Enzymatic Reaction Generation}
\label{app:synthetic_reaction_generation}

Stage-1 pretraining uses a large synthetic reaction corpus constructed
with a template-based generation framework following
RSGPT~\citep{deng2025rsgpt}.
The overall pipeline is
\begin{equation}
\begin{split}
    \text{enzymatic reactions}
    &\rightarrow
    \text{reaction templates},
    \\
    \text{molecular collection}
    &\rightarrow
    \text{BRICS fragments},
    \\
    \text{templates}+\text{fragments}
    &\rightarrow
    \text{candidate reactions}
    \rightarrow
    \text{filtered synthetic corpus}.
\end{split}
\label{eq:synthetic_data_pipeline}
\end{equation}

The following sections describe each stage in detail.

\subsection{Reaction Template Construction}
\label{app:template_extraction}

A reaction template represents a transferable local transformation
rather than a complete reaction instance.
Starting from an atom-mapped reaction, we compare the reactant and
product sides to identify atoms and bonds whose local environments
change.
The transformation together with the required neighboring structural
context defines a reaction template.

Conceptually, a template $T$ with $K$ reactant patterns can be written
as
\begin{equation}
    T:
    \left(
        Q_1,Q_2,\ldots,Q_K
    \right)
    \rightarrow Q_P,
    \label{eq:reaction_template}
\end{equation}
where $Q_i$ denotes a reactant-side substructure query and $Q_P$
encodes the corresponding product-side transformation.

The query patterns preserve the local chemical constraints required for
template application, including relevant atom and bond properties.
Templates are normalized and duplicate transformation rules are merged
while retaining their provenance to the source enzymatic reactions.

Each template additionally retains its association with the catalytic
annotations of its source reactions when available.
Such an inherited EC annotation describes the reaction category from
which the template was obtained.
It should not be interpreted as experimental evidence that a particular
enzyme catalyzes every synthetic reaction generated from that template.

\subsection{Candidate Substrate Pool}
\label{app:synthetic_fragment_pool}

To expose each transformation to a larger molecular context, we
construct a candidate substrate pool following the fragment-based
strategy used in RSGPT~\citep{deng2025rsgpt}.
Molecules from the upstream molecular collection are decomposed using
the BRICS fragmentation algorithm~\citep{degen2008art}.

BRICS identifies chemically meaningful bonds that can be disconnected
to produce smaller molecular fragments.
After fragmentation, attachment points are converted into chemically
valid terminated structures, yielding independently parseable
fragments.
The resulting structures are standardized and deduplicated before
template matching.

Let
\begin{equation}
    \mathcal{F}
    =
    \{f_1,f_2,\ldots,f_N\}
\end{equation}
denote the resulting fragment pool.
These fragments provide candidate substrate contexts; BRICS connection
points themselves do not define the enzymatic reaction center.

\subsection{Template--Fragment Matching}
\label{app:template_fragment_matching}

For every reactant pattern $Q_i$ of template $T$, we identify fragments
whose molecular structures satisfy the corresponding substructure
query.
We define the candidate set
\begin{equation}
    \mathcal{F}(T,i)
    =
    \left\{
        f\in\mathcal{F}
        \mid
        f \models Q_i
    \right\},
    \label{eq:template_fragment_pool}
\end{equation}
where $f\models Q_i$ indicates that fragment $f$ contains a valid match
to the template's $i$-th reactant pattern.

For a template containing $K$ reactant components, the candidate
combination space is therefore
\begin{equation}
    \mathcal{C}(T)
    =
    \mathcal{F}(T,1)
    \times
    \mathcal{F}(T,2)
    \times \cdots \times
    \mathcal{F}(T,K).
    \label{eq:template_combination_space}
\end{equation}

This combinatorial construction is the main source of data scaling:
a single experimentally observed transformation can be instantiated
with many compatible substrate combinations.
Importantly, matching is based on the structural constraints encoded by
the reaction template rather than molecular similarity alone.

\subsection{Template-Based Reaction Generation}
\label{app:template_application}

For each compatible reactant combination
\begin{equation}
    \mathbf{f}
    =
    (f_1,\ldots,f_K)
    \in
    \mathcal{C}(T),
\end{equation}
we apply template $T$ in the forward direction to obtain one or more
candidate products:
\begin{equation}
    \mathcal{P}_{T}(\mathbf{f})
    =
    T(f_1,\ldots,f_K).
\end{equation}

Each successful template application produces a candidate reaction
\begin{equation}
    f_1.f_2.\cdots.f_K
    \;\longrightarrow\;
    p,
    \qquad
    p\in\mathcal{P}_{T}(\mathbf{f}).
\end{equation}

A reactant combination may fail to produce a valid product, and a
single combination may yield several products because of multiple
matching sites.
Such outcomes are handled before the reaction is admitted to the
pretraining corpus.

For traceability, generated reactions retain their template provenance
and, where available, the catalytic annotation inherited from the
source enzymatic transformation.

\subsection{Reaction Filtering}
\label{app:synthetic_reaction_filtering}

Template application enforces a local reaction rule, but it does not
guarantee that every generated instance represents a plausible
biochemical reaction.
We therefore apply several filtering stages before pretraining.

\paragraph{Structural validity.}
Reactants and products must be parseable molecular structures and the
template application must produce a non-empty product.
Invalid molecular structures and failed template applications are
discarded.

\paragraph{Reaction consistency.}
We remove degenerate transformations and normalize component ordering
and molecular representations before reaction-level comparison.
Generated reactions are deduplicated after normalization while
retaining template and EC provenance when multiple templates produce
the same transformation.

\paragraph{Biochemical and thermodynamic filtering.}
We further filter the generated candidates using the biochemical
constraints and reaction-energy information available in our upstream
processing pipeline.
This step removes candidates that are inconsistent with the target
enzymatic reaction space and prioritizes thermodynamically plausible
transformations.


\paragraph{Benchmark isolation.}
Finally, synthetic pretraining reactions are filtered independently for
each \benchmark split.
We remove direct overlaps with held-out reactions and additionally
exclude records associated with templates that occur exclusively in
held-out matches.
Let $T_{\mathrm{train}}$, $T_{\mathrm{val}}$, and
$T_{\mathrm{test}}$ denote template sets associated with reactions
matched to each benchmark partition.
The excluded held-out template set is
\begin{equation}
    T_{\mathrm{held}}
    =
    \left(
        T_{\mathrm{val}}
        \cup
        T_{\mathrm{test}}
    \right)
    \setminus
    T_{\mathrm{train}}.
    \label{eq:heldout_template_filter}
\end{equation}
Synthetic records directly matching validation or test reactions, or
associated with a template in $T_{\mathrm{held}}$, are removed from
the corresponding pretraining corpus.

\subsection{From Synthetic Reactions to Model Inputs}
\label{app:synthetic_to_rsmiles}

After generation and filtering, the retained synthetic reactions enter
the same representation-processing pipeline used by \model.

We first obtain atom mappings with RXNMapper when a valid mapping is not
already available~\citep{schwaller2021rxnmapper}.
We then construct up to 20 root-aligned SMILES views for each reaction
using R-SMILES~\citep{zhong2022root}.
Forward and retrosynthetic views are generated according to their
respective prediction directions and stored together for downstream
multitask training.

This stage should be distinguished from synthetic reaction generation.
Template application increases the number of distinct reaction
instances,
\begin{equation}
    |\mathcal{D}_{\mathrm{reaction}}|,
\end{equation}
whereas R-SMILES increases the number of sequence realizations of each
instance,
\begin{equation}
    r
    \longrightarrow
    \{
        s_r^{(1)},\ldots,s_r^{(M)}
    \},
    \qquad M\leq20.
\end{equation}

Therefore, the approximately four million synthetic reactions used for
Stage-1 pretraining refer to reaction instances before counting their
alternative R-SMILES representations.

\subsection{Interpretation of Synthetic Reactions}
\label{app:synthetic_reaction_scope}

The generated corpus is intended as large-scale pretraining data rather
than a database of experimentally verified enzymatic reactions.
Each generated example satisfies an extracted local transformation rule
and the implemented filtering criteria, but template compatibility alone
does not establish experimental feasibility.

Similarly, an EC label inherited from a template identifies the
catalytic category associated with the source transformation.
It does not establish that a specific enzyme carrying that EC annotation
accepts the newly generated substrates.

The synthetic corpus is therefore used to expose \model~to a broader
enzymatic reaction space before the model is adapted to experimentally
observed \benchmark reactions during Stage-2 continued training.

\section{\model-Tiny results without constrained generation}

\begin{table*}[t]
\centering
\caption{\model-Tiny results without constrained generation on the three data splits.
We report Top-$k$ exact-match accuracy (\%).
These results use the standard unconstrained decoding protocol.}
\label{tab:model_without_constrained}
\scriptsize
\setlength{\tabcolsep}{3.2pt}

\begin{tabular}{llcccccccccccc}
\toprule
\multirow{2}{*}{\textbf{Task}} &
\multirow{2}{*}{\textbf{EC Input}} &
\multicolumn{4}{c}{\textbf{Reactant Split}} &
\multicolumn{4}{c}{\textbf{Product Split}} &
\multicolumn{4}{c}{\textbf{EC Split}} \\
\cmidrule(lr){3-6}\cmidrule(lr){7-10}\cmidrule(lr){11-14}
& & T1 & T3 & T5 & T10
& T1 & T3 & T5 & T10
& T1 & T3 & T5 & T10 \\
\midrule

\multirow{2}{*}{Forward}
& Without EC
& 73.066 & 84.611 & 88.048 & 90.739
& 50.959 & 66.275 & 71.393 & 77.085
& 31.100 & 45.170 & 53.236 & 59.947 \\

& With EC
& 75.221 & 86.778 & 89.725 & 92.474
& 55.825 & 69.463 & 74.041 & 78.811
& 34.564 & 50.240 & 56.340 & 64.058 \\

\midrule

\multirow{2}{*}{Retrosynthesis}
& Without EC
& 8.446 & 17.183 & 23.777 & 30.603
& 20.554 & 30.058 & 34.923 & 39.945
& 4.878 & 8.737 & 10.487 & 14.274 \\

& With EC
& 17.230 & 26.398 & 29.345 & 34.425
& 28.847 & 34.696 & 38.267 & 40.568
& 18.360 & 24.605 & 27.481 & 30.872 \\

\bottomrule
\end{tabular}
\end{table*}
\FloatBarrier
\section{Additional Experimental Results}
\FloatBarrier
\subsection{Retrosynthesis}
\label{sec:round_trip}

\begin{table*}[t]
\centering
\caption{
Round-trip coverage on three data splits.
We report RT Coverage@$k$ (\%), defined as the percentage of test reactions
for which at least one of the top-$k$ retrosynthesis candidates can be
mapped back to the target product by the forward oracle.
Methods marked with $^{*}$ are zero-shot methods and are evaluated directly using open-source checkpoints.
Methods are grouped by whether EC information is provided to the
retrosynthesis model.
Within each group and column, the best result is bold and the second-best result is underlined.
}
\label{tab:round_trip_coverage}
\scriptsize
\setlength{\tabcolsep}{3.2pt}

\begin{tabular}{lcccccccccccc}
\toprule
\multirow{2}{*}{\textbf{Method}} &
\multicolumn{4}{c}{\textbf{Reactant Split}} &
\multicolumn{4}{c}{\textbf{Product Split}} &
\multicolumn{4}{c}{\textbf{EC Split}} \\
\cmidrule(lr){2-5}
\cmidrule(lr){6-9}
\cmidrule(lr){10-13}
& C@1 & C@3 & C@5 & C@10
& C@1 & C@3 & C@5 & C@10
& C@1 & C@3 & C@5 & C@10 \\
\midrule

\multicolumn{13}{l}{\textit{Zero-shot}} \\
SciReasoner$^*$ (LLM)
& \textbf{31.163} & \textbf{45.853} & \textbf{52.260} & \textbf{61.125}
& \textbf{20.206} & \textbf{30.489} & \textbf{36.266} & \textbf{44.487}
& \textbf{17.953} & \textbf{29.782} & \textbf{36.937} & \textbf{46.045} \\

RSGPT$^*$
& \underline{14.247} & \underline{25.524} & \underline{29.555} & \underline{34.098}
& \underline{8.881} & \underline{14.861} & \underline{16.982} & \underline{19.427}
& \underline{8.198} & \underline{17.797} & \underline{21.201} & \underline{24.904} \\

R-SMILES$^*$
& 3.856 & 9.879 & 13.898 & 19.385
& 2.721 & 6.807 & 10.139 & 14.849
& 4.830 & 8.401 & 12.548 & 19.691 \\

\midrule

\multicolumn{13}{l}{\textit{Without EC Input}} \\
Graph2SMILES
& 38.269 & 52.551 & 58.504 & 63.781
& 28.667 & 34.552 & 37.140 & 42.174
& 25.443 & 34.744 & 38.698 & 43.313 \\

Retroformer
& 24.814 & 34.098 & 41.344 & 51.083
& 10.331 & 14.669 & 17.965 & 22.651
& 15.292 & 21.453 & 25.791 & 31.771 \\

MEGAN
& 37.710 & 53.763 & 60.252 & 66.577
& 30.513 & 44.451 & 50.803 & 57.083
& 21.213 & 35.031 & 41.587 & 49.269 \\

R-SMILES
& 29.928 & 42.661 & 47.029 & 52.575
& 15.832 & 28.164 & 35.031 & 42.581
& 10.930 & 22.016 & 28.092 & 37.452 \\

BioNavi
& 39.108 & 52.365 & 58.388 & 64.189
& 11.397 & 17.414 & 20.686 & 24.724
& 20.973 & 29.279 & 33.030 & 38.219 \\

G2Retro
& 31.139 & 51.549 & 58.854 & 67.754
& 27.205 & 40.448 & 49.029 & 58.293
& 22.699 & 39.070 & 46.069 & 56.843 \\

RSGPT
& 22.623 & 42.428 & 49.219 & 60.217
& 12.021 & 21.908 & 26.282 & 33.869
& 13.627 & 24.065 & 30.717 & 39.250 \\

\model
& \textbf{43.383} & \textbf{60.648} & \textbf{66.787} & \textbf{75.128}
& \textbf{41.107} & \textbf{54.602} & \textbf{60.954} & \textbf{69.607}
& \textbf{28.883} & \textbf{48.394} & \textbf{55.081} & \textbf{64.753} \\

\model~(MLCD)
& \underline{43.057} & \underline{58.586} & \underline{63.269} & \underline{70.282}
& \underline{38.255} & \underline{49.940} & \underline{56.520} & \underline{62.105}
& \underline{28.655} & \underline{47.711} & \underline{53.152} & \underline{60.990} \\

\midrule

\multicolumn{13}{l}{\textit{With EC Input}} \\
ECReact
& 44.851 & 60.135 & 65.284 & 70.701
& 28.152 & 36.601 & 41.407 & 47.291
& 32.862 & 43.349 & 48.370 & 55.633 \\

\model
& \textbf{53.308} & \textbf{66.880} & \textbf{72.064} & \textbf{76.887}
& \textbf{47.915} & \textbf{56.544} & \textbf{63.015} & \textbf{69.307}
& \textbf{42.006} & \textbf{53.428} & \textbf{58.221} & \textbf{65.964} \\

\model~(MLCD)
& \underline{50.233} & \underline{63.723} & \underline{67.719} & \underline{74.348}
& \underline{43.996} & \underline{53.248} & \underline{57.251} & \underline{62.668}
& \underline{38.231} & \underline{50.096} & \underline{53.955} & \underline{60.822} \\

\bottomrule
\end{tabular}
\end{table*}

\begin{table*}[t]
\centering
\caption{
Round-trip accuracy on three data splits.
We report RT Accuracy@$k$ (\%), defined as the fraction of the first $k$
retrosynthesis candidates that can be mapped back to the target product
by the forward oracle.
Methods marked with $^{*}$ are zero-shot methods and are evaluated directly using open-source checkpoints.
Methods are grouped by whether EC information is provided to the
retrosynthesis model.
Within each group and column, the best result is bold and the second-best result is underlined.
}
\label{tab:round_trip_accuracy}
\scriptsize
\setlength{\tabcolsep}{3.2pt}

\begin{tabular}{lcccccccccccc}
\toprule
\multirow{2}{*}{\textbf{Method}} &
\multicolumn{4}{c}{\textbf{Reactant Split}} &
\multicolumn{4}{c}{\textbf{Product Split}} &
\multicolumn{4}{c}{\textbf{EC Split}} \\
\cmidrule(lr){2-5}
\cmidrule(lr){6-9}
\cmidrule(lr){10-13}
& A@1 & A@3 & A@5 & A@10
& A@1 & A@3 & A@5 & A@10
& A@1 & A@3 & A@5 & A@10 \\
\midrule

\multicolumn{13}{l}{\textit{Zero-shot}} \\
SciReasoner$^*$ (LLM)
& \textbf{31.163} & \textbf{24.282} & \textbf{20.503} & \textbf{17.672}
& \textbf{20.206} & \textbf{14.533} & \textbf{12.081} & \textbf{9.340}
& \textbf{17.953} & \textbf{13.774} & \textbf{12.157} & \textbf{9.939} \\

RSGPT$^*$
& \underline{14.247} & \underline{12.950} & \underline{10.848} & \underline{7.780}
& \underline{8.881} & \underline{6.935} & \underline{5.297} & \underline{3.496}
& \underline{8.198} & \underline{7.674} & \underline{6.764} & \underline{4.715} \\

R-SMILES$^*$
& 3.856 & 4.042 & 4.268 & 3.929
& 2.721 & 2.621 & 2.812 & 2.641
& 4.830 & 3.356 & 3.490 & 3.476 \\

\midrule

\multicolumn{13}{l}{\textit{Without EC Input}} \\
Graph2SMILES
& 38.269 & 33.652 & 30.594 & 25.495
& 28.667 & 22.064 & 18.502 & 14.403
& 25.443 & 22.108 & 19.578 & 15.511 \\

Retroformer
& 24.814 & 22.930 & 22.260 & 21.553
& 10.331 & 9.416 & 8.758 & 7.296
& 15.292 & 14.446 & 13.320 & 11.956 \\

MEGAN
& 37.710 & 35.089 & 32.968 & 29.065
& 30.513 & 25.559 & 23.418 & \underline{20.150}
& 21.213 & 19.451 & 18.068 & 15.690 \\

R-SMILES
& 29.928 & 26.130 & 24.117 & 21.262
& 15.832 & 12.568 & 11.644 & 9.950
& 10.930 & 10.067 & 9.267 & 8.201 \\

BioNavi
& 39.108 & \underline{35.702} & \underline{33.979} & \underline{30.325}
& 11.397 & 9.368 & 8.205 & 6.510
& 20.973 & 18.333 & 16.285 & 13.424 \\

G2Retro
& 31.139 & 27.512 & 24.497 & 21.454
& 27.205 & 20.877 & 19.456 & 16.419
& 22.699 & 20.254 & 17.543 & 14.880 \\

RSGPT
& 22.623 & 20.857 & 20.315 & 18.113
& 12.021 & 9.907 & 8.411 & 6.874
& 13.627 & 12.672 & 11.663 & 9.884 \\

\model
& \textbf{43.383} & \textbf{37.508} & \textbf{34.434} & \textbf{30.494}
& \textbf{41.107} & \textbf{31.516} & \textbf{28.346} & \textbf{25.345}
& \textbf{28.883} & \textbf{25.803} & \textbf{23.027} & \textbf{20.152} \\

\model~(MLCD)
& \underline{43.057} & 35.038 & 31.379 & 27.340
& \underline{38.255} & \underline{27.744} & \underline{24.295} & 18.948
& \underline{28.655} & \underline{24.549} & \underline{21.026} & \underline{16.710} \\

\midrule

\multicolumn{13}{l}{\textit{With EC Input}} \\
ECReact
& 44.851 & \underline{42.280} & \textbf{39.814} & \textbf{35.216}
& 28.152 & 27.201 & \underline{26.515} & \underline{24.895}
& 32.862 & \textbf{32.882} & \textbf{31.853} & \textbf{30.128} \\

\model
& \textbf{53.308} & \textbf{43.480} & \underline{38.844} & \underline{34.287}
& \textbf{47.915} & \textbf{36.326} & \textbf{32.728} & \textbf{27.096}
& \textbf{42.006} & \underline{31.819} & \underline{28.032} & \underline{23.688} \\

\model~(MLCD)
& \underline{50.233} & 39.418 & 33.684 & 28.879
& \underline{43.996} & \underline{30.661} & 25.686 & 20.147
& \underline{38.231} & 26.127 & 22.090 & 17.338 \\

\bottomrule
\end{tabular}
\end{table*}

\FloatBarrier
\subsection{EC prediction}

\begin{table*}[t]
\centering
\caption{
EC-number prediction on the Reactant and Product splits.
Macro F1 (\%) is reported at EC levels 1--4.
}
\label{tab:ec_all_macro_f1}
\scriptsize
\setlength{\tabcolsep}{3.2pt}
\renewcommand{\arraystretch}{1.08}
\resizebox{\textwidth}{!}{%
\begin{tabular}{lcccccccc}
\toprule
\multirow{2}{*}{\textbf{Method}} &
\multicolumn{4}{c}{\textbf{Reactant Split}} &
\multicolumn{4}{c}{\textbf{Product Split}} \\
\cmidrule(lr){2-5}\cmidrule(lr){6-9}
& EC1 & EC2 & EC3 & EC4 & EC1 & EC2 & EC3 & EC4 \\
\midrule

\multicolumn{9}{l}{\textit{Zero-Shot}} \\

crxnfp~\citep{zhang2025crxnfp}
& $\underline{86.64}$ &
$71.17$ &
$57.28$ &
$27.35$ &
$68.45$ &
$55.85$ &
$46.12$ &
$21.12$ \\

DRFP~\citep{probst2022drfp}
& $85.75$ &
$\underline{72.80}$ &
$\underline{57.53}$ &
$\underline{31.94}$ &
$\underline{71.11}$ &
$\underline{60.03}$ &
$\underline{51.23}$ &
$\underline{26.15}$ \\

rxnfp~\citep{schwaller2021rxnfp}
& $83.71$ &
$70.56$ &
$56.36$ &
$28.46$ &
$66.90$ &
$54.32$ &
$45.66$ &
$21.11$ \\

\textbf{\model}$^{*}$
& $\mathbf{87.38}$ &
$\mathbf{73.05}$ &
$\mathbf{61.05}$ &
$\mathbf{33.31}$ &
$\mathbf{73.46}$ &
$\mathbf{61.14}$ &
$\mathbf{51.96}$ &
$\mathbf{26.90}$ \\

\midrule

\multicolumn{9}{l}{\textit{Task Specific Model}} \\

Theia~\citep{probst2023theia}
& $83.72 \pm 1.70$ &
$71.34 \pm 0.95$ &
$\mathbf{62.15} \pm 1.75$ &
$\mathbf{36.29} \pm 0.53$ &
$73.72 \pm 2.03$ &
$\mathbf{63.79} \pm 1.23$ &
$\underline{52.92} \pm 1.07$ &
$\mathbf{27.81} \pm 0.02$ \\

CLAIRE~\citep{zeng2025claire}
& $\mathbf{89.80} \pm 0.66$ &
$\mathbf{72.32} \pm 2.93$ &
$59.51 \pm 1.57$ &
$31.74 \pm 0.74$ &
$74.50 \pm 0.22$ &
$57.82 \pm 0.68$ &
$49.99 \pm 0.82$ &
$23.96 \pm 0.56$ \\

BEC-Pred~\citep{qian2024becpred}
& $85.35 \pm 6.25$ &
$61.58 \pm 1.10$ &
$42.64 \pm 0.59$ &
$28.13 \pm 0.43$ &
$\mathbf{76.70} \pm 1.49$ &
$60.23 \pm 1.03$ &
$43.30 \pm 0.89$ &
$21.18 \pm 0.27$ \\

\textbf{\model}
& $\underline{88.35} \pm 1.54$ &
$\underline{71.65} \pm 0.64$ &
$\underline{62.10} \pm 0.63$ &
$\underline{33.93} \pm 0.93$ &
$\underline{76.01} \pm 0.69$ &
$\underline{62.12} \pm 1.88$ &
$\mathbf{54.95} \pm 0.57$ &
$\underline{26.09} \pm 0.21$ \\

\bottomrule
\end{tabular}%
}

\vspace{2pt}
\begin{minipage}{\textwidth}
\footnotesize
{\model} denotes frozen \model~representations followed by a task-specific predictor.
Task-specific results are mean $\pm$ std over three seeds.
\end{minipage}
\end{table*}

\FloatBarrier
\section{Details of Molecule Library-Constrained Decoding}
\label{app:mlcd}

\subsection{implementation}

MLCD constrains autoregressive decoding with a prefix tree constructed
from the tokenized molecular representations in the predefined molecule
library.
Each trie node corresponds to a molecular token prefix, and its outgoing
edges specify the valid next tokens.
A molecule can terminate only when the current prefix reaches a complete
library entry.

For multi-molecule outputs, the trie is reset after each molecule
separator so that the constraint is applied independently to every
generated molecule.
Reaction-level control tokens and separators are generated normally.
MLCD therefore changes only the set of admissible molecular tokens;
sequence probabilities and beam ranking remain determined by \model.

Algorithm~\ref{alg:mlcd} summarizes the decoding procedure.

\begin{algorithm}[t]
\caption{Molecule Library-Constrained Decoding (MLCD)}
\label{alg:mlcd}
\begin{algorithmic}[1]
\Require Input $x$, model $p_\theta$, molecule trie $\mathcal{T}$,
beam size $B$
\Ensure Ranked output sequences

\State Initialize beam set $\mathcal{B}$ with the start token

\While{not all beams are finished}
    \State $\mathcal{B}_{\mathrm{new}} \gets \emptyset$

    \For{each partial sequence $y \in \mathcal{B}$}
        \If{$y$ is currently generating a molecule}
            \State $h \gets$ current molecular prefix
            \State $\mathcal{A} \gets$ valid next tokens from $\mathcal{T}$ given $h$
        \Else
            \State $\mathcal{A} \gets$ valid reaction-level tokens
        \EndIf

        \State Expand $y$ using tokens in $\mathcal{A}$
        \State Score expansions with $p_\theta(\cdot \mid x,y)$
    \EndFor

    \State Keep the top-$B$ scored expansions
    \State Update $\mathcal{B}$
\EndWhile

\State \Return completed sequences ranked by model score
\end{algorithmic}
\end{algorithm}

\subsection{Analysis of Molecule Library-Constrained Decoding}
\label{app:mlcd_analysis}

We provide additional analysis of Molecule Library-Constrained Decoding
(MLCD) for retrosynthesis.

\paragraph{Generation validity.}
We report the raw invalid rate, defined as the percentage of generated
molecular candidates that cannot be parsed as valid SMILES before final
filtering.
Missing Top-$k$ slots are not counted as invalid molecules.

\begin{table}[t]
\centering
\caption{
Raw invalid SMILES rate (\%) for retrosynthesis with and without MLCD.
Lower is better.
}
\label{tab:invalid_smiles}
\scriptsize
\setlength{\tabcolsep}{5pt}

\begin{tabular}{lccc}
\toprule
\textbf{Split} &
\textbf{EC} &
\textbf{Unconstrained} &
\textbf{MLCD} \\
\midrule
\multirow{2}{*}{Reactant}
& \checkmark & 2.277 & \textbf{0.000} \\
& --         & 1.599 & \textbf{0.000} \\
\midrule
\multirow{2}{*}{Product}
& \checkmark & 3.641 & \textbf{0.000} \\
& --         & 2.756 & \textbf{0.000} \\
\midrule
\multirow{2}{*}{EC}
& \checkmark & 3.086 & \textbf{0.000} \\
& --         & 2.210 & \textbf{0.000} \\
\bottomrule
\end{tabular}
\end{table}

As shown in Table~\ref{tab:invalid_smiles}, MLCD eliminates invalid raw
molecular candidates in all evaluated retrosynthesis settings.
This supports its role in constraining generation to structurally valid
molecular outputs.

\paragraph{Round-trip evaluation.}
For each retrosynthesis candidate, we apply a fixed forward oracle and
test whether the predicted product matches the target product.
We report RT Accuracy@$k$, the fraction of the first $k$ candidates that
map back to the target, and RT Coverage@$k$, the fraction of test
reactions for which at least one of the first $k$ candidates maps back
to the target.

Tables~\ref{tab:round_trip_accuracy} and
\ref{tab:round_trip_coverage} show that unconstrained decoding generally
achieves higher round-trip scores than MLCD.
Combined with the higher exact-match accuracy of MLCD, this indicates
that constrained decoding improves recovery of recorded benchmark
targets while reducing the diversity of alternative candidates retained
by unconstrained decoding.

\end{document}